\documentclass[]{article}
\usepackage{fullpage,amsthm}
\usepackage{latexsym,amssymb,amsmath}

\def\eps{\varepsilon}
\def\Z{{\mathbb Z}}

\def\C{{\mathbb C}}

\newcommand{\HSP}{\mbox{\rmfamily\textsc{Hidden Subgroup}}}
\newcommand{\HTP}{\mbox{\rmfamily\textsc{Hidden Translation}}}
\newcommand{\OP}{\mbox{\rmfamily\textsc{Translating Coset}}}
\newcommand{\OCP}{\OP}
\newcommand{\OCSP}{\mbox{\rmfamily\textsc{Translating Coset Superposition}}}
\newcommand{\SP}{\mbox{\rmfamily\textsc{Stabilizer}}}
\newcommand{\SSP}{\mbox{\rmfamily\textsc{Stabilizer Superposition}}}
\newcommand{\OSP}{\mbox{\rmfamily\textsc{Orbit Superposition}}}
\newcommand{\GSP}{\mbox{\rmfamily\textsc{Subgroup Superposition}}}
\newcommand{\Order}{\mathrm{O}}

\newtheorem{theorem}{Theorem}
\numberwithin{theorem}{section} %
\newtheorem{lemma}[theorem]{Lemma}
\newtheorem{proposition}[theorem]{Proposition}
\newtheorem{corollary}[theorem]{Corollary}
\newtheorem{definition}[theorem]{Definition}

\begin{document}
\title{Hidden Translation and Translating Coset in Quantum Computing%
\thanks{A preliminary version of this paper appeared in
\cite{fimss03}.\newline
Partially supported by 
the European Commission IST STREP projects Quantum Computer Science (QCS) 255961
and Quantum Algorithms (QALGO) 600700,
by the French ANR Blanc program under contract ANR-12-BS02-005 (RDAM project), 
and
by the Hungarian National Science Fund (OTKA), grants NK105645 and K77476.
Research at the Centre for Quantum Technologies is funded by the Singapore Ministry of Education 
and the National Research Foundation,
also through the Tier 3 Grant ``Random numbers from quantum processes".\newline
Part of the work of the last three authors was done
while visiting MSRI, Berkeley and part of the work
of the second author was done during visits
at the CQT in Singapore.}}
\date{}
\author{Katalin Friedl\thanks{Budapest University of Technology and Economics, Hungary 
}
\and
G\'abor Ivanyos\thanks{Institute for Computer Science and Control, Hungarian Academy of Sciences, Budapest, Hungary
}
\and
Fr\'ed\'eric Magniez\thanks{CNRS, LIAFA, Univ Paris Diderot, Sorbonne Paris-Cit\'e, Paris, France 
}
\and
Miklos Santha\footnotemark[4]\ \thanks{Centre for Quantum Technologies, National University of Singapore, Singapore
}\and
Pranab Sen\thanks{Tata Institute of Fundamental Research, Mumbai, India 
}}

\maketitle

\newcommand{\ket}[1]{{|{#1}\rangle}}
\newcommand{\qft}{\mathrm{QFT}}
\newcommand{\size}[1]{\lvert #1 \rvert}
\newcommand{\norm}[1]{\left\lVert #1 \right\rVert}
\newcommand{\abs}[1]{\lvert #1 \rvert}
\newcommand{\dist}{\mathsf{dist}}
\newcommand{\prob}{\mathop{\mathsf{Pr}}}
\newcommand{\range}{\mathop{\mathsf{Range}}}
\renewcommand{\max}{\mathop{\mathsf{Max}}}
\renewcommand{\min}{\mathop{\mathsf{Min}}}
\newcommand{\group}[1]{{\langle}#1{\rangle}}
\newcommand{\vect}[1]{\mathrm{Span}(#1)}
\newcommand{\lint}[1]{\lfloor #1 \rfloor}
\newcommand{\cint}[1]{\lceil #1 \rceil}
\newcommand{\poly}{\mathrm{poly}}
\newcommand{\polylog}{\mathrm{polylog}}

\newcommand{\per}{\mathrm{Per}}
\newcommand{\equalcoset}{\mathrm{Range}}
\newcommand{\enclength}{\ell}

\newsavebox{\fmbox}
\newenvironment{fmpage}[1]
     {\begin{lrbox}{\fmbox}\begin{minipage}{#1}}
     {\end{minipage}\end{lrbox}\fbox{\usebox{\fmbox}}}

\newenvironment{algo}
{\ttfamily\small\begin{quote}}
{\end{quote}}

\renewenvironment{quote}{%
  \smallskip\list{}{%
    \leftmargin.7cm   
    \rightmargin\leftmargin
  }
  \item\relax
}
{\endlist\smallskip}

\newcommand{\algof}[1]{\mbox{\tt {#1}}}

\newcommand{\AGS}{AbelianGS} 
\newcommand{\EATCS}{ElementaryAbelianTCS} 
\newcommand{\TF}{TranslationFinding} 
\newcommand{\AATCS}{ArbitraryAbelianTCS} 
\newcommand{\OS}{OS} 
\newcommand{\TCS}{TCS} 

\begin{abstract}
We give efficient quantum algorithms for the problems of
\HTP{} and \HSP{} in
a large class of non-abelian solvable groups including solvable groups
of constant exponent and of constant length derived series.
Our algorithms are recursive. For the base case, we
solve efficiently \HTP{} in $\Z_{p}^{n}$,
whenever $p$ is a fixed prime.
For the induction step, we introduce the problem \OP{}
generalizing both \HTP{} and \HSP{}, and prove a powerful
self-reducibility result: 
\OP{} in a finite solvable group $G$
is reducible to instances of \OP{} in $G/N$ and $N$, for
appropriate normal subgroups $N$ of $G$.
Our self-reducibility framework combined with Kuperberg's subexponential
quantum algorithm for solving \HTP{} in any abelian group, leads to 
subexponential quantum algorithms for \HTP{} and \HSP{} in any
solvable group.
\end{abstract}


\section{Introduction}
Quantum computing is an extremely active research
area (for introductions see e.g.~\cite{ksv02,aha98,pre98,nc00}).
Many of the superpolynomial speedups achieved by quantum
algorithms over their best known classical counterparts
have been in a group theoretical setting.
In this setting,
we are given a finite group $G$ and, besides the group operations,  
we also have at our disposal 
a function $f$ mapping  $G$ into a finite set.
The function $f$  can be queried via an oracle.
The time complexity of an algorithm  is measured 
by the overall running time 
including both the queries 
(counting a query as one step)
and the quantum and/or classical processing of these
queries.
The most important unifying problem of group theory for the purpose
of quantum algorithms has turned out to be \HSP{},
which can be cast in the following broad terms:
Let $H$ be a subgroup of $G$ 
such that $f$ is constant on each left coset of $H$
and distinct on different left cosets.
We say that $f$ {\em hides} the subgroup $H$.
The task is to determine the {\em hidden subgroup} $H$.

While no classical algorithm can solve this problem with polynomial
query complexity even if $G$ is abelian, the biggest success of
quantum computing until now is that it can be solved by a quantum
algorithm efficiently for any abelian $G$.
We will refer to this quantum algorithm as the standard algorithm
for \HSP{}.
The main tool for this
solution is Fourier sampling based on
the (approximate) quantum Fourier transform for abelian
groups which
can be efficiently implemented quantumly~\cite{kit95}.
Simon's xor-mask finding~\cite{sim97},
Shor's factorization
and discrete logarithm finding algorithms~\cite{sho97},
and Kitaev's algorithm~\cite{kit95} for the abelian stabilizer problem
are all special cases of this general solution. 
Quantum 
algorithms of Hallgren~\cite{hallgren07,hallgren05} and
Schmidt and Vollmer~\cite{sv05} computing class groups and 
unit groups of number fields, including the solution of Pell's 
equation also follow these lines. 

Finding an efficient algorithm for \HSP{} for non-abelian groups $G$
is considered to be
one of the most important challenges at present in quantum computing.
Besides its intrinsic mathematical interest, the importance of this
problem is enhanced by the fact that it contains as a special case
the graph isomorphism problem.
Unfortunately, although its query complexity is shown
to be polynomial by Ettinger, H{\o}yer and Knill \cite{ehk04},
non-abelian \HSP{} seems to be much more difficult
than the abelian case. Although considerable efforts were spent
on it in the last few years, only a small number of successes can be 
reported. They can be divided into two categories. 
The standard abelian Fourier sampling based
algorithm has been extended to some non-abelian groups
in \cite{rb98,hrt00,gsvv01,gavinsky04,mrrs04,dmr10} 
using the quantum Fourier 
transform over these (non-abelian) groups.
Although efficient quantum Fourier transform
implementations are known for several non-abelian 
groups~\cite{bea97,hoy97,prb99,mrr04}, the power of
the technique appears to be very limited.
In a different approach, \HSP{} was efficiently solved in the
context of specific non-abelian black-box groups~\cite{bs84,wat00}
by~\cite{ims01} without using the Fourier transform
on the group, and instead using Fourier transforms over abelian groups
only. Similarly, only abelian Fourier transforms were used 
by~\cite{ig04,bcd05,chkl06,iss07,iss08} to solve the hidden subgroup problem in some
specific kinds of non-abelian groups. See  
\cite{cd10} for a more detailed review of hidden subgroup algorithms
and related problems.

In face of the apparent hardness of \HSP{} in non-abelian groups,
a natural line of research is to address subproblems of \HSP{} which,
in some groups, capture the main difficulty of the original problem.
In a pioneering paper, Ettinger and H{\o}yer~\cite{eh00}, in the case
of dihedral groups, implicitly considered another paradigmatic 
group problem, 
\HTP{}. 
Here we are given two injective functions $f_{0}$ and $f_{1}$ from
a finite group $G$ to some finite set such that,
for some group element $u$, the equality
$f_{1}(xu)=f_{0}(x)$ holds for every $x$. The task is to 
find the {\em translation} $u$.
In fact, whenever $G$ is abelian, \HTP{} is an instance of \HSP{} in 
the semi-direct product $G\rtimes \Z_2$, where the
hiding function is $f(x,b)=f_{b}(x)$. The group action in 
$G\rtimes \Z_2$ is defined as 
$(x_1, b_1)\cdot (x_2, b_2) = (x_1 + (-1)^{b_1} x_2, b_1 \oplus b_2)$, 
where $+$ denotes the 
group operation in $G$ and $\oplus$ denotes the group operation in
$\Z_2$.
In $G \rtimes \Z_2$, $f$ hides the subgroup $H = \{(0,0), (u,1)\}$.
Actually, there is an efficient
quantum reduction in the other direction as well,
and the two problems are quantum polynomial time 
equivalent~\cite{eh00}.
A nice consequence of this equivalence is that instead of dealing 
with \HSP{} in the non-abelian 
group $G\rtimes \Z_2$, we can address \HTP{} in
the abelian group $G$.
Ettinger and H{\o}yer~\cite{eh00} have shown that \HTP{}
can be solved by a two-step procedure when $G=\Z_N$ is cyclic:
a polynomial number of 
Fourier samplings over the abelian group $\Z_N\times\Z_{2}$ 
followed by an exponential time classical stage without further queries.
The best known quantum algorithm for \HTP{}
in cyclic (and, in general abelian) groups
is Kuperberg's subexponential time method \cite{kup03}.
Its relation to certain
lattice problems investigated by Regev~\cite{reg04}
provides evidence that \HTP{} in cyclic groups
might be in fact difficult.

In a related work, van Dam, Hallgren and Ip~\cite{dhi03} gave 
efficient solutions
for three cases of what they call the hidden shift problem.
They also define another problem called the hidden coset problem which
generalizes hidden shift.
Their hidden coset problem can be viewed as a generalization of our \HTP{}
to not necessarily injective functions.
While their paper gives efficient quantum algorithms for some 
specific hidden coset problems, in general
the hidden coset problem is of exponential query complexity 
even in $\Z_2^n$.

Our first result (\textbf{Theorem~\ref{theorem:translation}}) 
is an efficient quantum algorithm for \HTP{} in the case of
elementary abelian $p$-groups, 
that is groups $\Z_p^n$, for any fixed prime number $p$.
The quantum part of our algorithm is the same as 
in Ettinger and H{\o}yer's~\cite{eh00}
procedure: it consists of performing
{Fourier sampling} over the abelian group $\Z_{p}^{n}\times\Z_{2}$.
But while their classical post processing requires exponential time, 
here we are able to recover classically the translation 
in polynomial time from
the samples.
It turns out that {Fourier sampling} produces vectors
$y$ non-orthogonal to the translation $u$, that is we obtain
linear inequations for the unknown $u$.
This is different from the situation in
the standard algorithm for the abelian \HSP{},
where only vectors orthogonal to the hidden subgroup are generated.
We show that, after a polynomial number of samplings, the
system of linear inequations has a unique solution with 
high probability,
which we are able to determine in deterministic polynomial time.
An immediate consequence of Theorem~\ref{theorem:translation}
is that \HSP{} in $\Z_p^n\rtimes \Z_2$ is efficiently solvable by
a quantum algorithm. 

To solve \HTP{} in other groups 
(which include abelian groups of constant exponent),
we embark in a radically new direction whose basic
idea is {\em self-reducibility}.
Since \HTP{} is not well-suited for this self-reducibility based
approach, we define a new paradigmatic group problem.
Notice that there is a natural combination of
\HTP{} with \HSP. This is the version of 
\HTP{} where the functions $f_0$ and $f_1$ are not
necessarily injective, but they are certain subgroup
hiding functions. Indeed, if $f_1$ hides a subgroup
$H$ and $f_0(x)=f_1(xu)$ for some $u\in G$ and for every $x\in G$ then
the set of all such elements $u$ form a right coset 
of $H$. (In the context of graph isomorphisms, the corresponding
problem would be determining all the bijections between the
vertex sets which are isomorphisms. This set is a coset of
the automorphism group of one of the graphs.) The self-reducibility
will be based on ``averaging" over normal subgroups so that
we actually get a problem over the factor group. We will
give an averaging procedure which
results in quantum superpositions. Therefore our new problem, called
\OP{}, 
is a combination of \HTP{} and \HSP{} where we have
quantum states as input\footnote{In the preliminary 
version~\cite{fimss03} of the present paper,
the problem \OP{} was called {\sc Orbit Coset}. This was due to the fact that
the problem is actually a constructive version of testing membership in orbits
of permutation groups.}.
\OP{} also involves quantum group actions, that is groups acting
on a finite set of mutually orthogonal quantum states.
Given two such states $\ket{\phi_0}$ and $\ket{\phi_1}$, 
the \OP{} problem consists of finding their {\em translating coset}, which is
defined to be
the stabilizer subgroup of $\ket{\phi_1}$ and a group element that 
maps $\ket{\phi_1}$ to $\ket{\phi_0}$.

It turns out that with a slight modification, our algorithm of 
Theorem~\ref{theorem:translation}
also works for \OP{} in $\Z_p^n$ whenever many copies
of the input states are given.
Moreover, we show that \OP{} has the following self-reducibility
property in any finite solvable group $G$: 
it is reducible to instances of \OP{} in $G/N$ and $N$, for
any normal subgroup $N\lhd G$ (\textbf{Theorem~\ref{op-op}}).
This is the first general self-reducibility result
for a problem subsuming \HSP{}.
The proof of the result
involves a new technique which is based upon constructing the uniform 
superposition of
the orbit of a given quantum state (\OSP{}).
The importance of generating specific superpositions
for solving important algorithmic problems has been observed before,
see for instance 
the paper of Aharonov and Ta-Shma~\cite{at03}. For example
generating the uniform superposition of all graphs isomorphic to a given
graph, which in fact is an instance of the \OSP{} problem of the
symmetric group $S_n$ acting on an $n$-vertex graph, 
would allow 
us to solve the graph isomorphism problem.
We show how \OSP{} is related 
to \OP{} (\textbf{Theorem~\ref{osp-op}}).
The self-reducibility of \OP{} combined with its solvability
for $\Z_p^n$ enables us to design an efficient quantum algorithm
for \OP{} in groups that we call
smoothly solvable groups (\textbf{Theorem~\ref{htp-hsp-thm}}).
These groups include solvable groups of constant exponent and
constant length derived series; in particular, unitriangular matrix groups
of constant dimension over finite fields of constant characteristic.
For the special case of \SP{} (i.e. \OP{} when 
$\ket{\phi_1} = \ket{\phi_0}$),
we obtain an efficient quantum algorithm for an even 
larger class of solvable groups viz. for solvable groups having
a smoothly solvable commutator subgroup 
(\textbf{Theorem~\ref{htp-hsp-thm}}).
As an immediate consequence, we get efficient quantum algorithms
for \HTP{} and \HSP{}
in the same groups as \OP{} and \SP{} respectively.
By combining our self-reducibility results above with 
Kuperberg's~\cite{kup03} subexponential time algorithm for \HTP{} in 
abelian groups, and using the fact that every solvable group $G$ has 
derived series of length $\Order(\log\log\size{G})$~\cite{glasby89},
we get subexponential time algorithms for \HTP{} and
\HSP{} in all solvable groups (\textbf{Theorem~\ref{kup-thm}}), and
quasi-polynomial time quantum algorithm for \HTP{} and \HSP{} in 
solvable groups
of constant exponent (\textbf{Theorem~\ref{loglog-thm}}).

\section{Preliminaries}
\subsection{Quantum computation background}
For a background on standard quantum computing, we refer the reader 
to~\cite{nc00,ksv02}.
We will consider problems whose inputs and outputs might be either 
classical or quantum.
Moreover most of our problems are promise problems where a part of the 
input is given by an oracle.
A {\em problem} is a relation ${\cal P}\subseteq I\times O$, where $I$ 
is the set of {\em inputs}, and $O$ the set of possible {\em outputs}.
For a family of functions $\cal F$,
an {\em oracle problem} is a family of relations 
$({\cal P}^f)_{f\in\cal F}$,
where $f$ ranges over the family $\cal F$.
The function $f$ is given by a quantum oracle, that is a unitary
matrix $U_f$ implementing the map $U_f\ket{x}\ket{0}=\ket{x}\ket{f(x)}$.

For any finite set $S$, we denote by $\ket{S}$
the uniform superposition of elements in $S$:
$\ket{S}=\frac{1}{\sqrt{|S|}}\sum_{x\in S}\ket{x}$, when $S\neq\emptyset$,
and $\ket{S}=\ket{\emptyset}$ when $S = \emptyset$, where 
$\ket{\emptyset}$ is a specific basis element.

A quantum algorithm is a quantum circuit consisting of
a succession of quantum gates.
Sometimes we describe quantum algorithms using intermediate measurements,
but they can always be replaced by unitary operations acting on the 
system plus ancilla qubits~\cite{akn98}.
The {\em output state} of the algorithm is defined to be the reduced 
state at the end
of the algorithm  of a special register of qubits, called the 
{\em output register}.
Namely, the output state of the algorithm is obtained by tracing out all 
but the qubits
of the output register at the end of the algorithm.

In this paper, we consider problems with many possible correct answers.
For example, an algorithm for {\HSP} is said to be correct if it outputs
any generating set for the hidden subgroup.
Therefore we say that a quantum algorithm or a unitary transformation 
{\em solves} a problem {\cal P} {\em with error $\eps$}, if for 
every input $i\in I$ it produces an output state whose trace 
distance is at most $\eps$
from some mixture over $\{o \in O : (i,o)\in {\cal P}\}$
(see e.g.~\cite{akn98} for a definition of trace distance).

The {\em time complexity} of an algorithm is the number of gates and 
oracle calls in the circuit.
For every problem, the {\em input size} is the number of classical or 
quantum bits of an input.
We say that a computational problem can be solved in {\em quantum 
time} $t(n)$ if there exists a quantum algorithm which solves 
the problem with bounded error
in time $t(n)$ where $n$ is the input size.

\subsection{Group theory background}
Recall that the {\em exponent} of a finite group is the least
common multiple of the order of its elements and an {\em elementary 
abelian group} is a group isomorphic to $\Z_p^n$ for some positive integer $n$
and for some prime $p$. Obviously, the exponent of $\Z_p^n$ is $p$.
Let $G$ be a finite group. If $X$ is a subset of $G$ then
$\group{X}$ denotes the subgroup of $G$ generated by $X$. 

\subsubsection{Black-box groups}
Our results concern groups represented
in the general framework of black-box groups~\cite{bs84,wat00}
with unique encoding.
In this model, the elements of a finite group $G$
are uniquely encoded by 
binary strings of length $\enclength$, and
the group operations are performed by an oracle (the black box).
The group is given in terms of a collection of generators
and oracle may actually define operations for
a potentially larger group.
We formally denote the encoding by a mapping $\mathsf{enc}$ from
$G$ to $\{0,1\}^\enclength$.
For quantum algorithms, the group oracle performs the group operations
reversibly; see \cite{wat00} for a detailed description.
The {\em encoding length} $\enclength$ 
has to be at least $\log\size{G}$, and 
is usually $\Order(\log\size{G})$.
We measure the running time of our algorithm in term of the input 
size $\enclength$. 
Several times in this paper we will be dealing with subgroups or factor groups 
of black-box groups
wherein we will still continue to measure the running time in terms of the
input length $\enclength$ for the original group $G$, since we continue 
to use the original encodings for the subgroup elements. 
But even in this case,
all the encoding lengths for all subgroups shall be $\Order(\log \size{G})$,
where $G$ is the original group.

We do assume in all our problems that the groups are input
by at most $\log\size{G}$ generators. This is legitimate as there
are several efficient methods, e.g., the quantum algorithms given in
\cite{wat01} or \cite{ims01} that produce at most
$\log\size{G}$ generators for a solvable black-box
group $G$, even if it is given by a larger set of generators. 
The input size corresponding to $G$ is set to $\enclength$,
instead of $\enclength\times \log\size{G}$, for convenience.

\subsubsection{Solvable groups}

A sequence 
$G_0\geq G_1\geq \ldots \geq G_{m}$
of subgroups is a {\em subnormal series}
of $G$ if each $G_i$ is a normal subgroup of $G_{i-1}$. 
We use the notation
$G_0\rhd G_1\rhd\ldots\rhd G_{m}$ for a subnormal series.
The {\em length} of such a series is $m$.

The group $G$ is a {\em solvable} group when
there exists a subnormal series
$G_0\rhd G_1\rhd\ldots\rhd G_{m}$
such that $G=G_0$, $G_{m}=\{1_G\}$
and the factors $G_i/G_{i+1}$ ($i=0,1,\ldots,m-1$) are abelian.

A natural way of constructing a subnormal series of the solvable group $G$
is to considered its derived subgroups.
For any group $H$, let us first define and denote
the {\em commutator subgroup} $H'$ of $H$
by $H'=\group{\{h^{-1}k^{-1}hk : h,k\in H\}}$.
Then the {\em derived subgroups} $G^{(i)}$ ($i=0,1,2,\ldots$) are defined 
by induction:
$G^{(0)}=G$; and the $(i+1)$th derived subgroup 
$G^{(i+1)}$ is defined as the commutator $(G^{(i)})'$  
of $G^{(i)}$.
All the subgroups $G^{(i)}$ are normal subgroups of $G^{(j)}$, for $0\leq j < i$.
Clearly the group $G$ is solvable if $G^{(d)}=\{1_G\}$ for some positive integer $d$
and the {\em derived length} of $G$ is the smallest such integer $d$. 
The {\em derived series} of a solvable group $G$ is 
the chain $G=G^{(0)}\rhd G^{(1)}\rhd\ldots\rhd G^{(d)}=\{1_G\}$.

In the case of an abelian group $G$, 
we have at our disposal~\cite{cm01} an efficiently computable isomorphism for 
the {\em cyclic decomposition}
$\theta: \Z_{p_1^{k_1}}\times\ldots\times\Z_{p_r^{k_r}} \rightarrow G$,
where $p_i^{k_i}$ are prime powers for primes $p_i$.
Whenever $G$ is solvable,
the decomposition of $G$ into its {derived series} 
can be computed by a classical randomized procedure~\cite{bcfls95}.

\subsubsection{Smooth groups}
We introduce a shorthand terminology for the specific class of solvable groups for which 
our method works in polynomial time.
We say that an abelian group $G$ is $(e,s)$-{\em smooth} if 
it has a subgroup $N$ of index at most $s$ with exponent
at most $e$. A subnormal series $G=G_0\rhd G_1\rhd\ldots\rhd G_m=\{1_G\}$ of 
a solvable group $G$ is
{\em $(e,s)$-smooth} if each factor group $G_{i-1}/G_i$ is $(e,s)$-smooth.
A solvable group $G$ is $(e,s)$-{\em smooth} 
if its derived series is $(e,s)$-smooth.


The methods of this paper will work in polynomial time for
$(e,s)$-smooth solvable groups $G$
with constant derived length and
with constant $e$ and $s=\poly(\log\size{G})$.
We introduce the shorthand terminology
{\em smoothly solvable} for such groups. 
Solvable groups having constant derived length and satisfying the 
property that the factors of the consecutive derived subgroups are
of exponent bounded by a constant are the most typical
examples of smoothly solvable groups. 
An example of such a solvable group is a
unitriangular matrix group of constant dimension over a finite field of 
constant characteristic. 

\subsubsection{Quantum Fourier sampling}
When $G$ is a finite abelian group, we identify with $G$ the set 
$\widehat{G}$ of characters
of $G$ via some fixed isomorphism $y\mapsto \chi_{y}$.
(For a group $G$ isomorphic to $\Z_{k}^n$, it is usual
to define $\chi_y(x)$ as $e^{\frac{2\pi i}{k}x\cdot y}$,
where $x\cdot y$ stands for the standard inner product
$\sum_{i=1}^n x_iy_i \pmod{k}$. Of course, this definition
requires -- and depends on -- an isomorphism of $G$ with $\Z_k^n$.)
The {\em orthogonal subgroup of $H\leq G$} is
defined as $H^{\perp}=\{y\in G : \forall h\in H, \chi_{y}(h)=1\}$.
The {\em quantum Fourier transform} over $G$ is 
the unitary transformation defined for every $x\in G$ by
$\qft_G \ket{x} = 
\tfrac{1}{\sqrt{\size{G}}}\sum_{y \in G} \chi_{y}(x)\ket{y}$.
For the sake of convenience, we will use the exact 
abelian quantum Fourier transform
in our algorithm. Actual implementations~\cite{kit95,mos04} introduce
only exponentially small errors. 

The following well known quantum Fourier sampling
algorithm will be used as a building block,
where $G$ is a finite abelian group, $S$ is a finite set
and $f:G\rightarrow S$ is given by a quantum oracle.
This algorithm is actually the main ingredient
for solving \HSP{} in abelian groups when the function $f$ hides
a subgroup $H \leq G$.
In that case, \algof{FourierSampling}${}^{f}(G)$ generates
the uniform distribution over $H^{\perp}$. In the algorithm,
$\ket{0}_{S}$ stands for an arbitrary but fixed element of $S$.
\pagebreak
\begin{algo}
\algof{FourierSampling}${}^{f}(G)$
\begin{enumerate}
\setlength{\itemsep}{0mm}
\item Create state $\frac{1}{\sqrt{\size{G}}}\sum_{x\in G}\ket{x}\ket{0}_{S}$.
\item Query function $f$.
\item Compute $\qft_{G}$ on first register.
\item Measure and output the first register.
\end{enumerate}
\end{algo}

A function $f:G\rightarrow \C^{S}$ is a {\em quantum function}
if, for every $x\in G$, the vector $\ket{f(x)}$ has unit norm,
and, for every $x,y\in G$, the vectors $\ket{f(x)}$ and $\ket{f(y)}$
are either the same or orthogonal.
We say that the quantum function $f$ is {\em given} by
a quantum oracle if we have at our disposal
a unitary transformation $U_f$ 
and its inverse $U_f^{-1}$
satisfying
$U_{f}\ket{x}\ket{0}=\ket{x}\ket{f(x)}$, for every $x\in G$.

\subsubsection{Order finding and generalized discrete logarithm}
We also assume for simplicity that have at our disposal a zero-error 
quantum
algorithm for computing the generalized discrete logarithm and for order 
finding.
Given a basis $h_1,h_2,\ldots,h_l$ of an abelian group $H$ and $h\in H$, 
the {\em generalized discrete logarithm} consists of finding nonnegative 
integers $\alpha_1,\alpha_2,\ldots,\alpha_l$
such that $h=h_1^{\alpha_1} h_2^{\alpha_2}\ldots h_l^{\alpha_l}$.
Given a group element $g$ in any group, {\em order finding} consists of
finding the smallest
positive integer $r$ such that $g^r$ is the identity element.

The actual implementations for period finding~\cite{sho97},
for the single basis element case of discrete logarithm~\cite{sho97}
and for the general case~\cite{ims01} introduce only exponentially small 
errors.
Note that for discrete logarithm, one can also use a generalization of 
the single basis element case by \cite{mos04} which runs without error
if one has access to single qubit rotation gates of arbitrary precision.

\subsection{The problems}
Here we define the problems we are dealing with.
Each problem is  parametrized by some fixed group, and potentially by some group action.
These are given, as we specified above, by oracles.
Some inputs, usually functions on the group, can also be given by oracles,
we will refer to them as {\em oracle inputs}.

Let $G$ be a finite group and let $f_0, f_1$ be 
two injective functions from $G$ to
some finite set $S$.
The couple of functions $(f_0,f_1)$ can equivalently be considered as a
single function $f : G \times \Z_2\rightarrow S$,
where by definition $f(x,b) = f_b(x)$.
We will use $f$ for $(f_{0},f_{1})$ when it is convenient in the
coming discussion.
We call an element $u\in G$ the {\em translation} of $f$ if
for every $x \in G$, we have $f_1(xu)=f_0(x)$.

\begin{quote}
\HTP$(G)$\\
\textit{Oracle input:} Two injective
functions $f_{0},f_{1}$ from $G$ to some finite set $S$ such that
$f=(f_{0},f_{1})$ has a translation $u\in G$.\\
\textit{Output:} $u$.
\end{quote}

For a finite group $G$ and a finite set $\Gamma$ of mutually orthogonal 
quantum states, we consider group actions of $G$ on $\Gamma$.
By definition, $\alpha: G\times \Gamma\rightarrow \Gamma$ is 
a {\em group action} if for every $x\in G$ the quantum function 
$\alpha_x:\ket{\phi}\mapsto \ket{\alpha(x,\ket{\phi})}$
is a permutation over $\Gamma$, such that
the map $x \mapsto \alpha_x$ is a homomorphism from $G$ to the
symmetric
group on $\Gamma$,
i.e.,
$\alpha_{1_G}$ is the identity map and 
$\alpha_x\circ \alpha_{y^{-1}}=\alpha_{xy^{-1}}$, 
for every $x,y\in G$.
We extend $\alpha$ linearly to superpositions over $\Gamma$.
(The conditon that $G$ permutes the orthonormal system
$\Gamma$ of states is essential; we do not consider general 
unitary actions $G$ on Hilbert spaces.) 
When the group action $\alpha$ is fixed,
we use the notation $\ket{x\cdot \phi}$ for 
the state $\ket{\alpha(x,\ket{\phi})}$.
Having a group action $\alpha$ at our disposal means having a quantum
oracle realizing the unitary transformation 
$\ket{x}\ket{\phi}\mapsto \ket{x}\ket{x\cdot \phi}$.
For any positive integer $t$, we denote by $\alpha^t$ the 
group action of $G$ on 
$\Gamma^t=\{\ket{\phi}^{\otimes t} : \ket{\phi}\in\Gamma\}$
defined by $\alpha^t(x,\ket{\phi}^{\otimes t})=
\ket{x\cdot\phi}^{\otimes t}$.
Observe that one can construct a quantum oracle for $\alpha^t$ using
$t$ queries to a quantum oracle for $\alpha$.
We need the notion of $\alpha^t$ for the following reason.
Below, we define problems involving group actions on quantum superpositions
where the input superpositions
cannot, in general, be cloned
(that is, it may be impossible to make
further copies of the input state from just one).
However, it will be possible to generate multiple independent copies of 
the input superpositions by a separate process before the start of our 
algorithm. Hence, in the interests of reducing the error of our algorithm, 
we start it off with several 
independent copies of the input superpositions. 
Our self-reducibility arguments will reduce the main problem into a bunch
of problems involving actions of smaller groups on quantum
superpositions. To solve each of these subproblems with small
error, we will require that the self-reduction process leave a sufficient
number of independent copies of the input superpositions
for a subproblem. This is easy to ensure since we start with a large
number of independent copies of the input superpositions to the original
problem. However, in order to achieve this goal, the self-reduction
process needs to act on several independent superpositions simultaneously
by the same group element.
The group action $\alpha^t$ captures this notion.
This notion will be crucial for our induction arguments.
Also note that the stabilizer and the translating coset, defined later, are 
the same for group actions $\alpha$ and $\alpha^t$.

The {\em stabilizer} of a state $\ket{\phi}\in \Gamma$ is the subgroup
$G_{\ket{\phi}}=\{x\in G : \ket{x\cdot \phi}=\ket{\phi}\}$.
Given $\ket{\phi}\in \Gamma$, the problem 
\SP$(G,\alpha,t)$
consists of finding $\Order(\log\size{G})$ generators for 
the subgroup $G_{\ket{\phi}}$, given $t$ copies of $\ket{\phi}$.
\begin{proposition}\label{sp-ab}
Let $G$ be a finite abelian group 
given as a black-box group with encoding length $\enclength$
and let $\alpha$ be 
a group action of $G$.
When $t=\Omega(\log(\size{G})\log(1/\eps))$,
then 
\SP$(G,\alpha,t)$ 
can be solved in quantum time 
$\mathrm{poly}(\enclength)\log(1/\eps)$
with error~$\eps$.
\end{proposition}
\begin{proof}
Let $\ket{\phi}^{\otimes t}$ be the input of \SP{}. 
Let $f$ be the quantum function on $G$ defined by
$\ket{f(x)}=\ket{x\cdot\phi}$, for every $x\in G$.
Observe that $f$ is an instance of the natural 
extension of \HSP{} to quantum functions
and it hides the stabilizer $G_{\ket{\phi}}$.

The algorithm for \SP{} is simply the standard algorithm 
for the abelian \HSP{}
with error $\eps$.
In the standard algorithm, every query is of the form $\ket{x}_G\ket{0}_S$.
We simulate the $i^{\text{th}}$ query $\ket{x}_G\ket{0}_S$
using the $i^{\text{th}}$ copy of $\ket{\phi}$.
The second register of the query
is swapped with $\ket{\phi}$, and then we let $x$ act on it.
We remark that the standard algorithm for abelian \HSP{} outputs
$\Order(\log \size{G})$ generators for the hidden subgroup.
\end{proof}

Note that in general,
the input superposition $\ket{\phi}^{\otimes t}$ gets destroyed
by the above algorithm.

The {\em orbit} of a state $\ket{\phi}\in \Gamma$ is the subset
$G(\ket{\phi})=\{\ket{x\cdot \phi}: x\in G\}$.
Define $\ket{G\cdot\varphi} =
\frac{1}{\sqrt{\size{G(\ket{\phi})}}}
\sum_{\ket{\varphi'}\in G(\ket{\varphi})}\ket{\varphi'}$. 
Equivalently, 
$\ket{G\cdot\phi} = \frac{1}{\sqrt{\size{G
}}}
\sum_{x\in G
}\ket{x\cdot\phi}$.
The {\em translating coset} of two states $\ket{\phi_0}$ 
and $\ket{\phi_1}$ of $\Gamma$
is the set $\{u\in G : \ket{u\cdot\phi_1}=\ket{\phi_0}\}$.
The translating coset of $\ket{\phi_0}$ and $\ket{\phi_1}$
is either empty or a left coset $uG_{\ket{\phi_1}}$ (or equivalently
a right coset $G_{\ket{\phi_0}}u$), for some
$u\in G$. If the latter case occurs, $\ket{\phi_0}$ and $\ket{\phi_1}$
have conjugate stabilizers: $G_{\ket{\phi_0}}=uG_{\ket{\phi_1}}u^{-1}$.
\OP{} is a generalization of \SP{}:

\begin{quote}
\OP$(G,\alpha,t)$\\
\textit{Input:} $t$ copies of two quantum states
$\ket{\phi_0},\ket{\phi_1}\in\Gamma$.\\
\textit{Output:}
\begin{itemize}
\item reject, if $G(\ket{\phi_0})\cap G(\ket{\phi_1})=\emptyset$;
\item $u\in G$ such that $\ket{u\cdot\phi_1}=\ket{\phi_0}$
and $\Order(\log\size{G})$ generators for $G_{\ket{\phi_1}}$, otherwise.
\end{itemize}
\end{quote}

For a function $f$ on $G$, 
define the superposition 
$\ket{f}=\frac{1}{\sqrt{\size{G}}}\sum_{g\in G}\ket{g}\ket{f(g)}$,
and for $x\in G$, 
define the 
function $x\cdot f:g\mapsto {f(gx)}$.
Let $\Gamma(f)=\{\ket{x\cdot f} : x\in G\}$.
Then a group element $x$ acts naturally on 
$\ket{f'}\in \Gamma(f)$ by mapping it to the 
superposition $\ket{x\cdot f'}$.
We call this group action the {\em translation action}.
The mapping $\ket{x}\ket{f'}\mapsto\ket{x}\ket{x\cdot f'}$
is realized by right multiplying the first register 
of $\ket{f'}$ by $x^{-1}$.
\begin{proposition}\label{htp-op}
Suppose $G$ is a finite group and let $t=\mathrm{poly}(\log\size{G})$.
Then \HSP$(G)$ (resp. \HTP$(G)$) 
can be solved with a call to
\SP$(G,\tau,t)$ (resp. \OP$(G,\tau,t)$),
where $\tau$ denotes the translation action.
\end{proposition}
\begin{proof}
Let $f$ be an instance of \HSP{}. Then the stabilizer 
of $\ket{f}$ is the group hidden by $f$.
Let $(f_0,f_1)$ be an instance of \HTP{}. Then the translating coset of
$\ket{f_0}$ and $\ket{f_1}$ is a singleton whose element is the
translation of $(f_0,f_1)$.
\end{proof}


\section{Hidden Translation in \lowercase{\large $\Z_p^n$}}
\label{zp}
In this section, we show that \HTP{}$(G)$
can be solved in 
quantum polynomial time 
in the special case
when $G=\Z_p^n$ for any fixed prime number $p>2$.
In this section we use the additive notation
for the group operation and 
$x\cdot y\in\Z_p$ stands for the standard inner
product for $x, y \in \Z_p^n$.
Since $\Z_2^n \rtimes Z_2$ is isomorphic to the abelian group 
$\Z_2^n \times Z_2$, one already has a quantum polynomial time algorithm
for \HTP{} in $\Z_2^n$ by reducing it to
\HSP{} in $\Z_2^{n+1}$ by the method of \cite{eh00}.

For the convenience of the reader we present our method using intermediate
measurements.
However, the measurements can always be eliminated, see~\cite{akn98},
giving a unitary and therefore reversible algorithm, possibly with errors.

The quantum part of our algorithm consists of performing
\algof{FourierSampling} over the abelian 
group $\Z_{p}^{n}\times\Z_{2}$.
It turns out that from the samples we will only use elements
of the form $(y,1)$. The important property of these elements $y$ 
is that they are {\em not} orthogonal to the hidden translation. 
Some properties of the distribution of the samples are stated 
for general abelian groups in the following lemma.
\begin{lemma}\label{sampling-lemma}
Let $G$ be a finite abelian group.
Let $f=(f_0,f_1)$, $f: G \times \Z_2 \rightarrow S$
be an instance of  \HTP$(G)$
having a translation $u\neq 0$.
Then \algof{FourierSampling}${}^f(G\times\Z_2)$ outputs an element
in $G\times\{1\}$ with probability $1/2$. Moreover, the probability
of sampling the element $(y,1)$ depends only on $\chi_y(u)$,
and is $0$ if and only if $y\in {u}^{\perp}$.
\end{lemma}
\begin{proof}
The state vector of 
$\mbox{\algof{FourierSampling}}^{f}(G \times \Z_2)$ before 
the final observation is 
$$
\frac{1}{2\size{G}}
\sum_{x\in G}
\sum_{y\in G} \sum_{c=0,1} \chi_y(x)
\big(1+({\scriptstyle -}1)^c\chi_y(u)\big)
\ket{y}\ket{c}\ket{f_0(x)}.
$$
Therefore the probability of sampling $(y,1)$ is proportional
to $|1-\chi_y(u)|^2$, 
whence the statement follows as $\chi_y(u)=1$ if and only if
$y\in u^\perp$ 
and 
$\sum_{y\in G}|1-\chi_y(u)|^2=
2\size{G}-2\sum_{y\in G}\chi_y(u)=2\size{G}$.
\end{proof}

When $G=\Z_p^n$, the value $\chi_y(u)=e^{\frac{2\pi i}{p}y\cdot u}$
depends only on the
inner product $y\cdot u$ over $\Z_p$, and $y\in u^\perp$
exactly when $y\cdot u=0$.
Therefore every $(y,1)$ generated satisfies $y\cdot u\neq 0$.
Thus the output distribution is different from the usual one obtained
for the abelian \HSP{}
where only vectors orthogonal to the hidden subgroup are generated.
We overcome the main obstacle, which is that we do not know 
the actual value of the inner product $y\cdot u$, 
by raising these inequations to the power $(p-1)$.
They become a system of polynomial equations of degree at most 
$(p - 1)$ 
since $a^{p-1}=1$ for every non-zero $a\in\Z_{p}$.
In general, solving systems of polynomial equations
over any finite field is NP-complete. 
But using the other special feature of our distribution,
which is that the probability of sampling $(y,1)$ depends
only on the inner product $y\cdot u$,
we are able to show that 
-- for fixed prime $p$ --
after a polynomial number
of samplings, 
our system of equations has a unique solution with constant
probability, and the solution can be found
in deterministic polynomial time.

To solve our system of polynomial equations,
we linearize it in the $(p{-}1)^{\text{th}}$ symmetric power of
$\Z_p^n$. We think of $\Z_p^n$ as an $n$-dimensional vector space
over $\Z_p$.
For a prime number $p$ and an integer $k \geq 0$, let 
$\Z_p^{(k)}[x_1,\ldots,x_n]$
be the $k^{\text{th}}$ symmetric power of $\Z_p^n$ which will 
be thought of as
the vector space, over the finite field $\Z_p$,
of homogeneous polynomials of 
degree $k$ in variables $x_1,\ldots,x_n$.
The monomials of degree $(p-1)$ form a basis of 
$\Z_p^{(p-1)}[x_1,\ldots,x_n]$,
whose dimension is therefore $\binom{n+p-2}{p-1}$, which
is polynomial in $n$ 
when $p$ is constant.
$\Z_p^{(1)}[x_1,\ldots,x_n]$ is isomorphic to $\Z_p^n$ as a vector
space. For two vectors
$Y_1, Y_2 \in \Z_p^{(p-1)}[x_1,\ldots,x_n]$, we denote their
standard inner product over the monomial basis by 
$Y_1 \cdot Y_2$.

For every $y = (a_1, \ldots , a_n) \in \Z_p^n$ and positive integer $k$, 
we define $y^{(k)} \in \Z_p^{(k)}[x_1,\ldots,x_n]$ as the polynomial 
$( \sum_{j=1}^n a_j x_j )^{k}$. 
For $y = (a_1, \ldots , a_n)$, $z = (b_1, \ldots, b_n)$ in $\Z_p^n$
and  positive integers $k, l$, we define the product
$y^{(k)} z^{(l)} \in \Z_p^{(k+l)}[x_1,\ldots,x_n]$ as the polynomial
$( \sum_{i=1}^n a_i x_i )^{k} (\sum_{j=1}^n b_j x_j )^{l}$. 
Now observe that if $u=(u_1, \ldots, u_n)$ is
the hidden translation vector , then the vector 
$u^{*} \in \Z_p^{(k)}[x_1,\ldots,x_n]$
which for every monomial $x_{1}^{e_1}\cdots x_{n}^{e_n}$ has
coordinate $u_{1}^{e_1}\cdots u_{n}^{e_n}$
satisfies $y^{(p-1)}\cdot u^{*}= (y \cdot u)^{p-1}$.
Therefore each linear inequation $y\cdot u\neq 0$ over $\Z_p^n$
will be transformed into the linear equation $y^{(p-1)}\cdot U=1$
over $\Z_p^{(p-1)}[x_1,\ldots,x_n]$, where $U$ is a 
$\dim \Z_p^{(p-1)}[x_1,\ldots,x_n]$-sized vector of unknowns.

We will see below that the vectors $y^{(p-1)}$ span the space
$\Z_p^{(p-1)}[x_1,\ldots,x_n]$ when
$y$ ranges over $\Z_p^n$.
Moreover, in what is the main part of our proof,
we show in Lemma~\ref{lemma:fraction} that
whenever the span of $y^{(p-1)}$ for the samples $y$ is
not $\Z_p^{(p-1)}[x_1,\ldots,x_n]$, our sampling process furnishes
with probability at least $1/p$ a vector $z \in \Z_p^n$ such that
$z^{(p-1)}$ is linearly
independent from the $y^{(p-1)}$ for the previously sampled $y$'s.
This immediately implies that if our sample size is of the order
of the dimension of $\Z_p^{(p-1)}[x_1,\ldots,x_n]$,
the span of $y^{(p-1)}$ for the samples $y$ is 
$\Z_p^{(p-1)}[x_1,\ldots,x_n]$ with high probability.
In that case,
the linear equations $y^{(p-1)}\cdot U=1$ have exactly one solution
which is $u^{*}$. From this unique solution one can 
easily recover a vector $v$ such that 
$v = au$ for some $0 < a < p$ (note that $v^{*}=u^{*}$).
Now $u$ can be found by checking
the $(p-1)$ possibilities.

The following combinatorial lemma is at the basis of the correctness
of our procedure.
\begin{lemma}[Line Lemma]
\label{lemma:line}
Let $y,z \in \Z_p^n$. For $1 \leq k \leq p-1$, define 
$L^{(k)}_{z,y} =  \{(z+ay)^{(k)}: 0 \leq a \leq k \}$.
Then for all $0 \leq l \leq k$, 
$z^{(l)} y^{(k-l)} \in \vect {L^{(k)}_{z,y}}$, where the span is taken with
$\Z_p$-coefficients.
\end{lemma}
\begin{proof}
Let $M^{(k)}_{z,y} = \{z^{(l)} y^{(k-l)} : 0 \leq l \leq k \}$.
Clearly, $\vect{L^{(k)}_{z,y}} \subseteq \vect{M^{(k)}_{z,y}}$.
We claim that the inverse inclusion is also true since 
the determinant of $L^{(k)}_{z,y}$ in $M^{(k)}_{z,y}$ is
non-zero in $\Z_p$. Indeed, it is
$\left(\prod_{l=0}^{k} {k \choose l}\right) V(0, 1, \ldots, k)$,
where $V$ denotes the Vandermonde determinant.
\end{proof}

\begin{proposition}
\label{prop:span-zp}
For $1 \leq k \leq p - 1$, 
$\Z_p^{(k)}[x_1,\ldots,x_n]$ is spanned by $y^{(k)}$ as $y$ ranges
over $\Z_p^n$.
\end{proposition}
\begin{proof}
We prove the proposition by induction on $k$.
The base case $k=1$ is trivial. Suppose the statement
holds for $k$, $1 \leq k < p-1$. 
Consider a monomial $M$ in $x_1,\ldots,x_n$ of degree $k+1$. If
$M = x_i^{k+1}$ for some $1 \leq i \leq n$, then $M$ trivially lies
in the span of $y^{(k+1)}$ as $y$ ranges over $\Z_p^n$.
Else, $M = x_i M'$ for some $1 \leq i \leq n$ and
degree $k$ monomial $M'$. Let $z \in \Z_p^n$. 
{From} Lemma~\ref{lemma:line}, we see that 
$x_i z^{(k)} \in \vect{\{(x_i+az)^{(k+1)}: 0 \leq a \leq k+1\}}$.
By induction hypothesis,
$M'$ lies in the span of $z^{(k)}$ as $z$ ranges over $\Z_p^n$.
Hence, $x_i M'$ lies in the span of $x_i z^{(k)}$ as $z$ ranges over 
$\Z_p^n$. Thus, 
$M\in\vect{\{(x_i+az)^{(k+1)}: 0\leq a\leq k+1, z\in \Z_p^n\}}$.
This shows that 
$\Z_p^{(k+1)}[x_1,\ldots,x_n]$ is spanned by $y^{(k+1)}$ as $y$ ranges
over $\Z_p^n$, 
completing the proof of the induction step and also that of
the proposition.
\end{proof}

We are now ready to prove the main lemma.
\begin{lemma}\label{lemma:fraction}
Let $u \in \Z_p^n$, $u \neq 0$ and
$W$ be a subspace of $\Z_p^{(p-1)}[x_1,\ldots,x_n]$. We set 
$R = \{y \in \Z_p^n : y^{(p-1)} \in W \}$.
For $k = 0, \ldots , p-1$, let 
$V_k = \{ y \in  \Z_p^n : y \cdot u = k \}$ and
$R_k = R \cap V_k$.
If $W \neq \Z_p^{(p-1)}[x_1,\ldots,x_n]$, then 
$\size{R_k} / \size{V_k} \leq (p-1)/p$ for $k = 1, \ldots , p-1$.
\end{lemma}
\begin{proof}
Observe that $R_k = \{ky : y \in R_1\}$ for $0 < k < p$.
Therefore the sets $R_k$, $0 < k < p$  have the same size. 
Observe also that the sets $V_k$, $0 \leq k < p$ have the same size, and
they partition $\Z_p^n$.
Hence the values
$\size{R_k} / \size{V_k}$ are the same for $0 < k < p$.

Since $W \neq \Z_p^{(p-1)}[x_1,\ldots,x_n]$,
Proposition~\ref{prop:span-zp} implies that 
$R \neq \Z_p^n$.
We consider two cases. In the first case, $V_0 \subseteq R$.
This implies that $R_1$ is a proper subset of $V_1$.
Choose any $y \in  V_1 \setminus R_1$. Then by Lemma~\ref{lemma:line},
in every coset of $\group{y}$ there is an element outside of $R$.
A coset of $\group{y}$ contains exactly one element from each
$V_k$, $k = 0, \ldots, p-1$. Hence $\cup_{k \neq 0} V_k$ is 
partitioned into equal parts, each part of size $(p-1)$, by 
intersecting with the cosets of $\group{y}$. In each part, there
is an element outside of $R$.
Therefore 
$\size{\cup_{k\neq 0} R_k}/\size{\cup_{k \neq 0} V_k}\leq (p-2)/(p-1)$.
Hence, $\size{R_k} / \size{V_k} \leq (p-2)/(p-1) < (p-1)/p$ for 
$k = 1, \ldots, p-1$, and the statement follows.

In the second case, $V_0 \not \subseteq R$. Therefore,
there is an element $y \in V_0 \setminus R_0$.
Then every $V_k$, $k = 0, \ldots, p-1$, is a union of cosets of 
$\group{y}$. Lemma~\ref{lemma:line} implies that 
every coset of $\group{y}$ contains an element outside of $R$.
This proves that
$\size{R_k} / \size{V_k} \leq (p-1)/p$ 
for $k = 0, \ldots, p-1$. This completes the proof of the lemma.
\end{proof}

We now specify the algorithm \algof{\TF} and
prove that, with high probability, it finds the hidden translation
in quantum polynomial time
when $p$ is constant.

\begin{algo}
\algof{\TF${}^{f}(\Z_p^n)$}
\begin{enumerate}
\setlength{\itemsep}{0mm}
\item[0.] If $f_{0}(0)=f_{1}(0)$ then output $0$.
\item $N\leftarrow 13 p {n+p-2 \choose p-1}$.
\item For $i=1,\ldots,N$ do \\
\null$\quad (z_{i}, b_i ) \leftarrow
\mbox{\algof{FourierSampling}}^{f}(\Z_p^n\times\Z_{2})$.
\item $\{y_1, \ldots , y_M \} \leftarrow \{z_i : b_i = 1 \}$.
\item For $i=1,\ldots,M$ do $Y_{i} \leftarrow y_i^{(p-1)}$.
\item Solve the system of linear equations\\
\null$\quad Y_1 \cdot U =1 , \ldots , Y_M \cdot U =1$.
\item If there are no solutions or more than one solution then abort.
\item Let $1 \leq j \leq n$ be such that the coefficient of 
$x_j^{p-1}$ is 1 in~$U$.
\item Let $v = (v_1, \ldots , v_n) \in \Z_p^n$ be such that
$v_j =1$ and $v_k$ is the coordinate of $x_k x_j^{p-2}$ in 
$U$ for $ k \neq j$.
\item Find $0 < a < p$ such that $f_0(0) = f_1(av)$.
\item Output $av$.
\end{enumerate}
\end{algo}

\begin{theorem}\label{theorem:translation}
For every prime number $p$, every integer $n\geq 1$,
and every function $f: \Z_p^n \times Z_2 \rightarrow S$ having a 
translation given via a quantum oracle,
algorithm {\rm\algof{\TF}}${}^{f}(\Z_p^n)$
aborts with probability less than $1/2$, and 
when it does not abort it outputs the translation of $f$.
The query complexity of the algorithm is
$\Order(p (n+p)^{p-1})$, and its time complexity is $(n+p)^{\Order(p)}$.
\end{theorem}
\begin{proof}
Because of Step~0 of the algorithm, we can suppose w.l.o.g.
that the translation $u$ of $f$ is non-zero.

If the algorithm does not abort, then $U=u^{*}$ is the 
unique solution of the system in Step~5.
When the coefficient of $x_j^{p-1}$ is $1$ in $U$, then $u_j \neq 0$.
Also, $u_k = u_j v_k$ for every $k$. 
Thus, $u = u_j v$ and $u$ is found in Step~9 for $a = u_j$.

{From} Lemma~\ref{sampling-lemma}, we see that 
the probability that the algorithm
$\mbox{\rm\algof{FourierSampling}}^{f}(\Z_p^n \times \Z_2)$ 
outputs $(y,1)$ for some $y$ is $1/2$. 
Therefore the expected value of $M$ is $N/2$,
and $M < N/3$ with probability at most $e^{-N/18}< 1/4$ because of 
Chernoff's bound.
If the system $Y_1, \ldots , Y_M$ has full rank, then it has a unique
solution.
By Lemmas~\ref{sampling-lemma} and~\ref{lemma:fraction}, 
the expected number of linear equations that guarantee
that the system has full rank is at most $p\binom{n+p-2}{p-1}$.
Since $N/3 > 4p\binom{n+p-2}{p-1}$, by Markov's inequality,
the solution $U$ is unique with probability at least $3/4$.
Thus, the total probability of aborting is less than $1/2$.
\end{proof}

\begin{corollary}\label{htp-cor0}
Let $p$ be a prime. Then the problem of
\HTP$(\Z_p^n)$ can be solved in quantum time 
$(n+p)^{\Order(p)}\log(1/\eps)$ with error $\eps$ using 
$t = \Theta(p (n+p)^{p-1} \log(1/\eps))$ accesses to the oracles
for $f_0, f_1$.
\end{corollary}
\begin{proof}
We perform two modifications in the
algorithm \algof{\TF}.
First, to get error $\eps$, the integer $N$ is multiplied 
by $\Order(\log(1/\eps))$.
Moreover, we assumed in the algorithm that there is an 
oracle for $f=(f_0,f_1)$,
which was used to choose $f_b$ knowing $b$.
This is not possible in general when
$f_0$ and $f_1$ are given by two distinct oracles.
Therefore we replace the oracle access
$\ket{x}\ket{b}\ket{0}_S\mapsto\ket{x}\ket{b}\ket{f_b(x)}_S$ by 
$$\ket{x}\ket{b}\ket{0}_S\ket{0}_S\mapsto
\ket{x}\ket{b}\ket{f_b(x)}_S\ket{f_{1-b}(-x)}_S.
$$
This type of quantum oracle corresponds to the function
$f'=(f'_0,f'_1)$, where $f'_0(x)=(f_0(x),f_1(x))$
and $f'_1(x)=(f_1(x),f_0(-x))$. Obviously, $f'_0$ is
injective and $f'_0(x)=f'_1(x+u)$. 
We can apply Theorem~\ref{theorem:translation}
in this new setting.

Let us now show how to simulate this new oracle access.
{From} $\ket{x}\ket{b}\ket{0}_S\ket{0}_S$ we compute
$\ket{{\scriptstyle (-1)}^b x}\ket{b}\ket{0}_S\ket{0}_S$, and then
we call $f_0$ and get 
$\ket{{\scriptstyle (-1)}^b x}
\ket{b}\ket{f_0({\scriptstyle (-1)}^b x)}_S\ket{0}_S$.
We multiply the first register by $({-}1)$ and call $f_1$ which gives
$\ket{{\scriptstyle (-1)}^{b+1} x}
\ket{b}\ket{f_0({\scriptstyle (-1)}^b x)}_S
\ket{f_1({\scriptstyle (-1)}^{b+1} x)}_S$. 
Finally, we multiply the first
register by ${\scriptstyle (-1)}^{b+1}$, and swap the 
last two registers when $b=1$.
\end{proof}

As there is a quantum reduction from \HSP{}
in $\Z_p^n \rtimes \Z_2$ to \HTP{} 
in $\Z_p^n$ by the method of \cite{eh00}, we obtain the following 
corollary.
\begin{corollary}\label{cor:hsp}
Let $p$ be a fixed prime. Then
\HSP{}$(\Z_{p}^{n}\rtimes\Z_{2})$
can be solved in quantum 
time $\poly(n)$.
\end{corollary}

The algorithm \algof{\TF} can also be extended to
solve \OP{} in $\Z_p^n$.
\begin{corollary}\label{op-zp}
Let $p$ be a prime.
Let $\alpha$ be a group action of $\Z_p^n$.
When $t=\Omega(p(n+p)^{p-1}\log(1/\eps))$, \OP$(\Z_p^n,\alpha,t)$ 
can be solved in quantum time $(n+p)^{\Order(p)}\log(1/\eps)$
with error $\eps$.
\end{corollary}
\begin{proof}
Let the input of the \OP{}$(\Z_p^n,\alpha,t)$ be
$(\ket{\phi_0}^{\otimes t},\ket{\phi_1}^{\otimes t})$.
We can suppose w.l.o.g. that the stabilizers of 
$\ket{\phi_0}$ and $\ket{\phi_1}$
are trivial. Indeed the stabilizers can be computed 
by Proposition~\ref{sp-ab}.
If they are different then the algorithm obviously has 
to reject, otherwise
we work in the factor group $\Z_p^n/G_{\ket{\phi_0}}\cong\Z_p^{n'}$, 
for some $n'\leq n$. 
To be more specific, we can compute a ($\Z_p$-basis for) a subgroup
$G_1$ of $\Z_p^n$ which is a direct complement of $G_{\ket{\phi_0}}$
by augmenting a basis for $G_{\ket{\phi_0}}$ to a basis
of $\Z_p^n$ and we can actually work with $G_1$ in place of $G$.

For  $b=0,1$, let $f_b$ be the injective quantum 
function on $G$ defined by
$\ket{f_b(x)}=\ket{x\cdot\phi_b}$, for every $x\in G$.
If the translating coset of $(\ket{\phi_0},\ket{\phi_1})$
is empty, then $f_0$ and $f_1$ have distinct ranges. 
Otherwise the translating coset of $(\ket{\phi_0},\ket{\phi_1})$ 
is a singleton $\{u\}$, and $(f_0,f_1)$ have the translation $u$.

The algorithm for \OP{} on input 
$(\ket{\phi_0}^{\otimes t}, \ket{\phi_1}^{\otimes t})$
is the algorithm \algof{\TF} on input 
$(f_0, f_1)$ with
a few modifications described below.
The oracle access to $(f_0, f_1)$ is modified in the same way as
in 
Corollary~\ref{htp-cor0}.
We simulate the $i^{\text{th}}$ 
query $\ket{x}\ket{b}\ket{0}_S\ket{0}_S$
using the $i^{\text{th}}$ copy of $\ket{\phi_0}\ket{\phi_1}$.
The two registers $\ket{0}_S\ket{0}_S$ 
are swapped with $\ket{\phi_b}\ket{\phi_{1-b}}$,
and then we let act $x$ on $\ket{\phi_b}$ and 
$(-x)$ on $\ket{\phi_{1-b}}$.

The equality tests in steps~0 and~9 are replaced by the 
swap test~\cite{bcww01,gc01} iterated $\Order(\log(1/\eps))$ times.
Finally, $N$ is multiplied by $\Order(\log(1/\eps))$, and
the algorithm rejects whenever the algorithm 
\algof{\TF} aborts
or there is no solution in step~9.
\end{proof}

\section{Translating Coset in solvable groups}

\subsection{Preparation}

\subsubsection{Quantization of the problems}
Let $G$ be a black-box group with unique encoding
and let $\alpha$ be a group action on $\Gamma$.

We now describe quantum analogues of problems with classical outcomes,
as unitary transformations whose output are basically uniform superposition
on the possible classical outcomes.

We will give quantum circuit implementations for the new problems.
A quantum circuit has both {\em input/output} registers and {\em ancilla} registers.
The latter ones are initialized to some default value, usually a $0$-string,
that we denote $\ket{0}$. 
We will explicitly mention when we consider a different default value.
We identify a quantum circuit with the unitary transformation it defines.

Let $U$ be a unitary transformation.
A quantum circuit $C$ {\em implements $U$} if
$C=U\otimes \mathrm{Id}$, where the tensor product is between input/output registers and ancilla registers.
Most often, our unitary transformations will be only partially specified and our quantum circuits will
only approximately implement them.  This motivates the following generalization of implementation.

A {\em partial unitary} $U$ is a transformation defined on a subset $\mathcal{S}$ of a Hilbert space $H$,
such that there exists a unitary transformation $V$ on $H$ which coincides with $U$ on $\mathcal{S}$.
A quantum circuit $C$ {\em implements $U$ on $\mathcal{S}$ with error $\eps$}
if 
$C (\ket{\psi}\otimes\ket{0})$ and 
$U\ket{\psi}\otimes\ket{0}$ are of trace distance at most $\eps$,
for every $\ket{\psi}\in\mathcal{S}$.
We will omit $\eps$ when $\eps=0$, and also  $\mathcal{S}$ when it is understood from the context.

Given a circuit that implements a unitary $U$, one can design a circuit with same size that
implements the unitary $U^{-1}$ by applying backward the circuit for $U$, where each gate is replaced by its inverse. Therefore in our model, the complexity for implementing a unitary transformation
or its inverse is the same.
Thus we will say that a circuit uses as {\em black boxes $t$ implementations of $U$}
whenever it uses $t$ gates $U$ or $U^{-1}$.

Our notion of implementation does not allow any garbage in the computation: at the end of the computation
the ancilla registers must come back to their initial default value, potentially approximately.
It is not always true for a quantum circuit, even if it computes the desired outcomes.
In that case we will say that the computation is {\em with garbage}.
Nonetheless, when a quantum circuit computes a classical function without error, we can assume that the computation 
is without garbage using the standard cleaning method: run the circuit $C$, XOR the output in a new register 
(initialized to the $0$-string), undo the circuit by running $C^{-1}$.
In such a situation, we will therefore always assume that we have at our disposal
such a circuit without garbage.

\begin{definition}
Let $g_1,g_2,\ldots,g_k\in G$ and $H=\group{g_1,g_2,\ldots,g_k}$.
\GSP${}(G,g_1,g_2,\ldots,g_k)$ is a partial unitary transformation that
maps state $\ket{1_G}$ to state $\ket{H}$.
\end{definition}


In the following description of a quantum circuit, we write in boldface 
the input registers of the circuit, whereas fresh registers are in regular font.
The output registers are {\em a priori} the same as the input registers.
We also assume for simplicity that we have at our disposal a zero-error quantum
algorithm for computing the generalized discrete logarithm and for order finding.
The actual implementations for the single basis element case~\cite{sho97}
and for the general case~\cite{ims01} introduce only exponentially small errors.
Note that one can also use a generalization for the single basis element case of~\cite{mos04}
which is without error.
We also note that the heart of the circuit is not to compute $H$ from the  
generators, but rather to create the superposition over $H$ by  
uncomputing the discrete log.
\begin{algo}
\algof{\AGS{}${}(G,g_1,g_2,\ldots,g_k
)$}\\
{\em Hypothesis:} $H=\group{g_1,g_2,\ldots,g_l}$ is abelian.\\
Input: {\boldmath$\ket{1_G}$}
\begin{enumerate}
\item \label{basis} Compute a basis $h_1,h_2,\ldots,h_l$ such that
$\group{h_1}\times\group{h_2}\times\ldots
\times\group{h_l}=H$, and the respective orders $r_j$ of $h_j$.
\item Compute in a fresh register the superposition
$$\sum_{0\leq a_j<r_j}\ket{a_1,a_2,\ldots,a_l}\mbox{\boldmath$\ket{1_G}$}$$
\item\label{expo} Perform fast exponentiation $h_j^{a_j}$ in fresh register:
$$\sum_{0\leq a_j<r_j}\ket{a_1,a_2,\ldots,a_l}\ket{h_1^{a_1},h_2^{a_2},\ldots,h_l^{a_l}}\mbox{\boldmath$\ket{1_G}$}$$
\item Multiply $1_G$ by all the $h_j^{a_j}$:
$$\sum_{0\leq a_j<r_j}\ket{a_1,a_2,\ldots,a_l}\ket{h_1^{a_1},h_2^{a_2},\ldots,h_l^{a_l}}
\mbox{\boldmath$\ket{h_1^{a_1}h_2^{a_2}\ldots h_l^{a_l}}$}$$
\item Undo Step~\ref{expo}.
$$\sum_{0\leq a_j<r_j}\ket{a_1,a_2,\ldots,a_l}
\mbox{\boldmath$\ket{h_1^{a_1}h_2^{a_2}\ldots h_l^{a_l}}$}$$
\item Undo the computation of the generalized discrete logarithm of the group elements 
$h_1^{a_1}h_2^{a_2}\ldots h_l^{a_l}$ in the basis $(h_1,h_2,\ldots,h_l)$:
$$\sum_{0\leq a_j<r_j}
\mbox{\boldmath$\ket{h_1^{a_1}h_2^{a_2}\ldots h_l^{a_l}}$}=\mbox{\boldmath$\ket{H}$}$$
\item Undo Step~\ref{basis}.
\end{enumerate}
\end{algo}

\begin{theorem}\label{agsp}
Let $G$ be a black-box group with unique encoding of length $\enclength$.
Let $g_1,g_2,\ldots,g_k\in G$ be generators of an abelian subgroup $H$.
Then \algof{\AGS${}(G,g_1,g_2,\ldots,g_k)$} implements
\GSP${}(G,g_1,g_2,\ldots,g_k)$  in quantum time $\mathrm{poly}(k\ell)$.
\end{theorem}
\begin{proof}
Since the description of the algorithm is clear,
the proof consists in checking that all the tasks involved in \algof{\AGS${}(G,g_1,g_2,\ldots,g_k 
)$} can be done in the requested complexity.

The main potential difficulty is for Step~\ref{basis}.
This step can be done in quantum time $\mathrm{poly}(k\ell)$
using the method of~\cite{cm01}, without error since we assume that we can do quantum Fourier transform 
without error on abelian groups.
\end{proof}

For solvable groups, we consider the following extension which produces the
required superposition, but with garbage.
\begin{theorem}[\cite{wat01}]\label{sgsp}
Let $G$ be a black-box group with unique encoding of length $\enclength$.
Given generators $g_1,g_2,\ldots,g_k\in G$ of a solvable subgroup $H$,
the state $\ket{H}$ can be produced with error $\eps$ and with garbage
in quantum time $\mathrm{poly}(k\ell)\log (1/\eps)$.
\end{theorem}

Now we define the quantized versions
of \OCP{} and \SP{}. 
(These are descriptions of certain unitary transformations.)
Recall that if $T$ is empty then
$\ket{T}=\ket{\emptyset}$, 
where $\ket{\emptyset}$ is a specific basis element.

\begin{definition}
\OCSP$(G,\alpha,t)$ is the partial unitary transformation that
maps state $\ket{\phi_0}^{\otimes t}\ket{\phi_1}^{\otimes t}\ket{1_G}$ to state
$\ket{\phi_0}^{\otimes t}\ket{\phi_1}^{\otimes t}\ket{T}$,
where $T=\{u\in G:\ket{u\cdot \phi_1}=\ket{\phi_0}\}$.
\SSP{} is the special case of \OCSP{} with $\ket{\phi_1}=\ket{\phi_0}$.
\end{definition}

In general $O(\log\size{G}\log\frac{1}{\eps})$ 
copies of the coset superposition $\ket{T}$ are sufficient to determine $T$ 
classically with error probability $\eps$. To see this,
assume that 
we have state $\ket{T}^{\otimes s}$.
We then multiply the contents of the second, third, etc.~register
by the inverse of the group element in the first register. Then the first register
will contain an element representing the coset while in the remaining
register there are elements of the stabilizer subgroup which,
if $s$ is large enough, will contain a system of generators
with high probability.

\begin{algo}
\algof{\EATCS$(G,\alpha,t)$}\\
{\em Hypothesis}: $G\cong\Z_p^n$\\
Input: {\boldmath$\ket{\phi_0}^{\otimes t}\ket{\phi_1}^{\otimes t}\ket{1_G}$}
\begin{enumerate}
\item\label{ea} Apply the algorithm of Corollary~\ref{op-zp} on the first $2t$ input registers,
using a fresh register for the computation:
$$
\sum_{u\in G, X \in G^{\leq \log\size{G}}}
\alpha_{u,X}\ket{u,X}\ket{\theta_{u,X}}
\mbox{\boldmath$\ket{1_G}$},
$$
where $\ket{u,X}$ denotes the output of the algorithm of Corollary~\ref{op-zp},
and $\ket{\theta_{u,X}}$ denotes the other remaining registers.
\footnote{The sum is over all elements $u\in G$ and all list $X$ consisting
of at most $\log\size G$ elements of $G$. If there were no errors, 
$\alpha_{u,X}$ would be zero for pairs $(u,X)$ which do not describe 
the coset translating $\phi_0$ to $\phi_1$. Due to errors of 
the algorithm, some of such coefficients can be nonzero, although 
very small.}

\item Apply \algof{\AGS{}${}(G,X)$} 
to the last input register:
$$
\sum_{u\in G, X \in G^{\leq \log\size{G}}}
\alpha_{u,X}\ket{u,X}\ket{\theta_{u,X}}
\mbox{\boldmath$\ket{\group{X}}$}
$$
\item Left multiply the last input register by $u$:
$$
\sum_{u\in G, X \in G^{\leq \log\size{G}}}
\alpha_{u,X}\ket{u,X}\ket{\theta_{u,X}}
\mbox{\boldmath$\ket{u\group{X}}$}
$$
\item Undo Step~\ref{ea}.
\end{enumerate}
\end{algo}

\begin{corollary}\label{eaocsp}
Let $G\cong\Z_p^n$ 
be a black-box group with unique encoding of length $\enclength$.
Let $\alpha$ be a group action of $G$ and let $t=\Omega(p(n+p)^{p-1}\log(1/\eps))$
be a positive integer.
Then \algof{\EATCS$(G,\alpha,t)$}
implements \OCSP$(G,\alpha,t)$
with error $\eps$
in quantum time 
$\ell^{\Order(1)}(n+p)^{\Order(p)}\log(1/\eps)$.
\end{corollary}
\begin{proof}
In the first step of the algorithm, $X$ denotes a set of generators for $G_\ket{\phi_1}$
and $u$ a group element such that $\ket{u\cdot\phi_1}=\ket{\phi_0}$.
When no solution exists, we simply request the algorithm of  Corollary~\ref{op-zp}
to set $X=\emptyset$, and let $u$ be any group element, instead of rejecting.

Let $v$ be a fixed group element such that $\ket{v\cdot\phi_1}=\ket{\phi_0}$.
Because of the choice of the parameters
and by Corollary~\ref{op-zp},
the states $\ket{u\group{X}}$ and $\ket{v G_{\ket{\phi_1}}}$ 
are of trace distance at most $\eps$.
This implies that the final state of the algorithm is 
of trace distance at most $\eps$
from the following state without garbage:
$\ket{\phi_0}^{\otimes t}
\ket{\phi_1}^{\otimes t} \ket{vG_{\ket{\phi_1}}}.
$
\end{proof}

For an arbitrary abelian group $G$, we can modify procedure
\algof{\EATCS$(G,\alpha,t)$}
by replacing the algorithm
of Corollary~\ref{op-zp} with an adapted version of
Kuperberg's subexponential method
(see Theorem~7.1 of \cite{kup03}) to solve \OP. (We only need
modifications to Kuperberg's algorithm
like the ones to \algof{\TF} described in the proof
of Corollary~\ref{op-zp}:
We use the ``conditionally swapped pairs of functions" 
trick presented in the proof of 
Corollary~\ref{htp-cor0} and simulate the oracle with
input quantum states.) 
Let us call the resulting procedure 
\algof{\AATCS$(G,\alpha,t)$}.
We obtain the following.

\begin{corollary}\label{aaocsp}
Let $G$ 
be a black-box abelian group with unique encoding of length $\enclength$.
Let $\alpha$ be a group action of $G$ and let 
$t=2^{\Omega(\sqrt{\log\size{G}}})$
be a positive integer.
Then \algof{\AATCS$(G,\alpha,t)$}
implements \OCSP$(G,\alpha,t)$
with error $\eps$
in quantum time 
$\ell^{\Order(1)}2^{\Order({\sqrt{\log\size{G}}})}\log(1/\eps)$.
\end{corollary}

\subsubsection{Compatible encodings}
We will apply recursion into 
factor groups of solvable groups. Therefore we need an efficient procedure to design
a unique encoding for these factor groups.
Moreover, for the purpose of our algorithm we will require this encoding to be compatible with the
original encoding of the group in the following sense.
\begin{definition}\label{def-ce}
Let $G$ be a black-box group with unique encoding $\mathsf{enc}$ of length $\ell$.
Let $N$ be a normal subgroup of $G$.
A unique encoding $\mathsf{enc}_N$ for $G/N$ is {\em compatible} with $\mathsf{enc}$
if:
\begin{enumerate}
\item for every $x\in G$, there is $y\in xN$ such that $\mathsf{enc}_N(xN)=\mathsf{enc}(y)$,
\item 
 the partial unitary $\ket{\mathsf{enc}(x)}\ket{0}\mapsto\ket{\mathsf{enc}(x)}\ket{\mathsf{enc}_N(xN)}$,
 where $x\in G$,
 can be implemented in quantum time $\mathrm{poly}(\enclength)$.
\end{enumerate}
\end{definition}
Note that if $G$ has encoding length $\enclength$, then a compatible encoding for $G/N$ also
has encoding length $\enclength$.

{From} now on, we assume for simplicity
that we have at our disposal a multiple $r$
of $\size{G}$ such that $r=\Order(\ell)$. This multiple is given or computed once
for a group, and we keep the same value for all its subgroups.
This assumption is reasonable since for solvable groups the cardinality of $G$ can
be computed in time $\mathrm{poly}(\ell)$~\cite{wat01}.

In the following theorem, 
we assume for simplicity that have at our disposal a zero-error quantum
algorithm for computing the generalized discrete logarithm  and for order finding.
\begin{theorem}
\label{compatible-thm}
Let $G$ be a black-box solvable group with unique encoding $\mathsf{enc}$ of length $\ell$.
Let $N$ be a normal subgroup of $G$ 
such that $G/N$ is abelian. 
Assume that $\Order(\ell)$ copies of $\ket{N}$  are given.
There exists a unique encoding $\mathsf{enc}_N$ for $G/N$
such that:
\begin{enumerate}
\item A set of generators for $G/N$, whose size is at most $\log\size{G/N}$, 
can be computed in quantum time  $\mathrm{poly}(\enclength)$.
\item Group operations over $G/N$ using encoding $\mathsf{enc}_N$
can be computed in quantum time $\mathrm{poly}(\enclength)$.
\item $\mathsf{enc}_N$ is compatible with $\mathsf{enc}$.
\end{enumerate}
\end{theorem}
Note that even if all the tasks (1) and (2) will use as ancilla several copies of $\ket{N}$,
these copies are always restored at the end of the computations.
Indeed, since the outcomes of tasks (1) and (2) are classical, 
one can XOR their value in a fresh register, and reverse the procedure in order to be garbage free and 
restore the used copies of~$\ket{N}$.
\begin{proof}
Let $g_1,g_2,\ldots,g_k\in G$ be the generators defining $G$, where $k=\Order(\ell)$,
and let $r$ be a known multiple of $\size{G}$ such that $\log r=\Order(\ell)$.
The cosets $g_1N,g_2N,\ldots,g_kN$ are generators of $G/N$.
We now show how to learn the structure of the abelian group $G/N$, 
and in particular how to extract a subset of at most $\log\abs{G/N}$ generators.

Following the approach of~\cite{ims01}, we consider extensions of
the quantum algorithms for computing the generalized discrete logarithm  and for order finding
to functions having quantum ranges. More precisely for order finding,
the function $a\in\Z_r \mapsto \ket{x^aN}$ hides
the subgroup $\Z_{r_x}$, where $r_x$ is the order of $xN$.
This function is encoded by the partial unitary map $\ket{a}\ket{N}\mapsto \ket{a}\ket{x^aN}$,
which admits a $\mathrm{poly}(\enclength)$ size circuit, since it can be implemented
using $\Order(\log r)$ group operations. 
Thus the algorithm requires as many copies of $\ket{N}$ as the number
of function evaluations, that is $\Order(\ell)$.
Similarly, given $x\in G$ and $y\in x^a N$, for some unknown 
$0\leq a <\size{G/N}$, one can compute $a$ using $\Order(\ell)$ copies of $\ket{N}$.

More generally, one can learn the structure of $G/N$ as in~\cite{cm01}
using $\Order(\ell)$ copies of $\ket{N}$ and the unitary
$$\ket{a_1,a_2,\ldots,a_k}\ket{N}\mapsto
\ket{a_1,a_2,\ldots,a_k}\ket{g_1^{a_1}g_2^{a_2}\ldots g_k^{a_k}N},$$
where $a_i\in\Z_r$ ,
and the group elements $g_i$ are implicitly encoded using $\mathsf{enc}$.
Given the structure of $G/N$, we are able to find the
lexicographically smallest non-redundant 
subset of generators for $G/N$ from $g_1N,g_2N,\ldots,g_kN$
by throwing out $g_i$ if it is contained in the subgroup of $G$ generated
by $g_1,\ldots,g_{i-1}$ and $N$. Without loss of generality
we can assume that this set is $g_1N,g_2N,\ldots,g_jN$.
By  non-redundancy, we must have $j\leq \log\abs{G/N}$.
This full construction of generators $g_1N,g_2N,\ldots,g_jN$ 
can be done in quantum time $\mathrm{poly}(\ell)$, and therefore condition~(1) is satisfied. 

For every $i=1,2,\ldots,j$, let $l_i$ be the least positive integer such that 
$g_i^{l_i}\in\group{N,g_1,g_2,\ldots,g_{i-1}}$. Then we can define our 
compatible encoding by
$$\mathsf{enc}_N(xN)=\mathsf{enc}(g_1^{a_1}g_2^{a_2}\ldots g_j^{a_j}),\quad
\text{where  $xN=g_1^{a_1}g_2^{a_2}\ldots g_j^{a_j} N$,
and $0\leq a_i<l_i$}.$$
Since the exponents $a_i$ are uniquely defined, 
the encoding is unique and satisfies condition~(1) of the definition of compatible encodings
(Definition~\ref{def-ce}).
In order to satisfy the conditions of compatible encodings, and therefore condition~(3) of the theorem, 
we show how to compute in quantum time $\mathrm{poly}(\ell)$
$\mathsf{enc}_N(xN)$ from $\mathsf{enc}(x)$.
Again we follow the approach of~\cite{ims01}.
Consider the unitary $$\ket{b,b_1,b_2,\ldots,b_j}\ket{N}\mapsto
\ket{b,b_1,b_2,\ldots,b_j}\ket{x^{-b}g_1^{b_1}g_2^{b_2}\ldots g_j^{b_j}N}.$$
This unitary hides a subgroup $H$ of $\Z_r\times\Z_{l_1}\times\cdots\times\Z_{j}$ 
generated by a generator of type
$u=(1,a_1,a_2,\ldots,a_j)$, where $xN=g_1^{a_1}g_2^{a_2}\ldots g_j^{a_j} N$. 
Therefore $\mathsf{enc}_N(xN)=\mathsf{enc}(g_1^{a_1}g_2^{a_2}\ldots g_j^{a_j})$.
The subgroup $H$, and therefore the generator $u$ of this particular form, 
can be found in quantum time 
$\mathrm{poly}(\ell)$ since this is the solution of \HSP{} 
for abelian groups extended 
to functions having quantum ranges~\cite{ims01}.

Finally, condition~(2) is easily satisfied.
Indeed, by the compatibility of our encoding, group operations over $G/N$
can be simulated by one call to the group oracle for $G$. 
Then the result $\mathsf{enc}(x)$, for some $x\in G$, has to be converted to $\mathsf{enc}_N(xN)$, using the above procedure.
\end{proof}

\subsection{Orbit Superposition}

In this section, we show that 
computing the uniform superposition of the orbit of a given state
 is reducible to instances of \OCSP{}.
In the following definition,
we denote by $\ket{G\cdot\varphi}$
the state $\frac{1}{\sqrt{\size{G(\ket{\phi})}}}
\sum_{\ket{\phi'}\in G(\ket{\phi})}\ket{\phi'}^{\otimes s}$,
where $\ket{\varphi}=\ket{\phi}^{\otimes s}$.
 
 \begin{definition}
Let $\ket{\phi}\in\Gamma$. Let $s$ be a positive integer
and $\ket{\varphi}=\ket{\phi}^{\otimes s}$.
\OSP$(G,\alpha,s)$
is the partial unitary transformation that
maps state $\ket{\varphi}\ket{\varphi}\ket{G}$ to
$\ket{G\cdot \varphi}\ket{\varphi}\ket{1_G}$.
\end{definition}
Then the following algorithm implements \OSP{}.
\begin{algo}
\algof{\OS$(G,\alpha,s)$}\\
Input: {\boldmath$\ket{\phi}^{\otimes 2s}\ket{G}$}
\begin{enumerate}
\item Apply the group element in 3rd register to the first $s$ registers:
$$\sum_{x\in G} \mbox{\boldmath$\ket{x\cdot \phi}^{\otimes s}\ket{\phi}^{\otimes s}\ket{x}$}
=\sum_{\ket{\varphi'}\in G(\ket{\varphi})} \mbox{\boldmath$\ket{\varphi'}\ket{\varphi}
\ket{xG_{\ket{\phi}}}$},
$$
where $\ket{\varphi}=\ket{\phi}^{\otimes s}$
\item Perform the inverse of 
\OCSP$(G,\alpha,s)$\\
(which maps $\ket{x\cdot\phi}^{\otimes s}\ket{\phi}^{\otimes s}\ket{1_G}$ to 
$\ket{x\cdot\phi}^{\otimes s}\ket{\phi}^{\otimes s}\ket{xG_{\ket{\phi}}}$):
$$\mbox{\boldmath$\ket{G\cdot \varphi}\ket{\varphi}\ket{1_G}$}
$$
\end{enumerate}
\end{algo}

\begin{theorem}\label{osp-op}
Let $G$ be a black-box group with unique encoding of 
length $\enclength$
and let $\alpha$ be a group action on $\Gamma$.
Let $\ket{\phi}\in\Gamma$. Let $s$ be a positive integer
 and $\ket{\varphi}=\ket{\phi}^{\otimes s}$.
\algof{\OS$(G,\alpha,s)$} implements
\OSP$(G,\alpha,s)$
using as a black box
one implementation of \OCSP$(G,\alpha,s)$
and quantum time $\mathrm{poly}(\enclength s)$ for the remaining computation.
\end{theorem}

\subsection{Translating Coset self-reducibility in solvable groups}
The purpose of this section is to prove Theorem~\ref{op-op}
stating the reducibility of  \OP{} in some solvable group $G$ to
\OP{} in proper normal subgroups and factors of $G$ under some conditions.
Given a group action $\alpha$ of $G$ on a finite set $\Gamma$
of mutually orthogonal quantum states, we define for every proper normal
subgroup $N\lhd G$ the group action $\alpha_N$ of $G/N$ on 
$\{\ket{N\cdot\phi} : \ket{\phi}\in\Gamma\}$ by
$\alpha_N(xN,\ket{N\cdot\phi})=\ket{x\cdot (N\cdot\phi)}$, for every 
$x\in G$ and $\ket{\phi}\in\Gamma$. Note that this action is 
independent of the chosen coset representative $x$, 
it only depends on
the coset $xN$.

For a group action like $\alpha^s$ on 
$\Gamma^s$
the group action $(\alpha^s)_N$ will act on states such as
$\ket{N\cdot \varphi}=\frac{1}{\sqrt{\size{N(\ket{\phi})}}}
\sum_{\ket{\phi'}\in N(\ket{\phi})}\ket{\phi'}^{\otimes s}$,
where $\ket{\phi}\in\Gamma$ and $\ket{\varphi}=\ket{\phi}^{\otimes s}$.

Note that, the use of our notion of compatible encodings
allows us to treat the oracle for $\alpha$ as an oracle for $\alpha_N$. 

In the following algorithm, we implicitly use the encoding $\mathsf{enc}$ of $G$
for its elements $z\in G$, and a compatible encoding $\mathsf{enc}_N$
for $G/N$ (given by Theorem~\ref{compatible-thm}) for its cosets $zN\in G/N$.
Last, for a subset $S\subseteq G$, the notation $S/N$ is the following subset of $G/N$:
$S/N=\{xN : x\in S\}.$
In particular, for any subgroup $H\leq G$, we have
$uHN/N=\{uhN : hN\in HN/N\}$.

\begin{algo}
\algof{\TCS$(G,N,\alpha,s(t+1))$}\\
{\em Hypothesis}: $N\lhd G$ with compatible encoding for $G/N$\\
Input: {\boldmath$\ket{\phi_0}^{\otimes s(t+1)}\ket{\phi_1}^{\otimes s(t+1)}
\ket{1_G}$}\\
Ancilla: $\ket{N}^{\otimes 2t}\ket{0}$
\begin{enumerate}
\item\label{osn} Perform $t$ times \algof{\OS$(N,\alpha,s)$} on blocks
$\ket{\phi_0}^{\otimes s}\ket{N}$\\
and then $t$ times on blocks $\ket{\phi_1}^{\otimes s}\ket{N}$:
$$\mbox{\boldmath
$\ket{N\cdot \varphi_0}^{\otimes t}\ket{\phi_0}^{\otimes s}
\ket{N\cdot \varphi_1}^{\otimes t}\ket{\phi_1}^{\otimes s}\ket{1_G}$}
\ket{1_G}^{\otimes 2t}\ket{0},
$$
where $\ket{\varphi_0}=\ket{\phi_0}^{\otimes s}$ and $\ket{\varphi_1}=\ket{\phi_1}^{\otimes s}$
\item XOR the compatible encoding of $1_{G/N}$ into the ancilla register $\ket{0}$:
$$\mbox{\boldmath
$\ket{N\cdot \varphi_0}^{\otimes t}\ket{\phi_0}^{\otimes s}
\ket{N\cdot \varphi_1}^{\otimes t}\ket{\phi_1}^{\otimes s}\ket{1_G}$}
\ket{1_G}^{\otimes 2t}\ket{1_{G/N}}
$$
\item Perform  \OCSP$(G/N,(\alpha^s)_N,t)$ on
$\ket{N\cdot \varphi_0}^{\otimes t}\ket{N\cdot \varphi_1}^{\otimes t}\ket{1_{G/N}}$:
$$\mbox{\boldmath$\ket{N\cdot \varphi_0}^{\otimes t}\ket{\phi_0}^{\otimes s}
\ket{N\cdot \varphi_1}^{\otimes t}\ket{\phi_1}^{\otimes s}\ket{1_G}$}
\ket{1_G}^{\otimes 2t}\ket{uHN/N},
$$
where $H=G_\ket{\phi_1}$ and $\ket{u\cdot\phi_1}=\ket{\phi_0}$, if there is any; 
$$\mbox{\boldmath$\ket{N\cdot \varphi_0}^{\otimes t}\ket{\phi_0}^{\otimes s}
\ket{N\cdot \varphi_1}^{\otimes t}\ket{\phi_1}^{\otimes s}\ket{1_G}$}
\ket{1_G}^{\otimes 2t}\ket{\emptyset},
$$
otherwise
\item Undo Step~\ref{osn}:
$$\mbox{\boldmath$\ket{\phi_0}^{\otimes s(t+1)}
\ket{\phi_1}^{\otimes s(t+1)}\ket{1_G}$}
\ket{N}^{\otimes 2t}\ket{uHN/N},
$$
or
$$\mbox{\boldmath$\ket{\phi_0}^{\otimes s(t+1)}
\ket{\phi_1}^{\otimes s(t+1)}\ket{1_G}$}
\ket{N}^{\otimes 2t}\ket{\emptyset}.
$$
In the second case, Stop the algorithm here
\item Perform $s$ applications of inverse of group element in the 
last register to registers $\ket{\phi_0}$ 
(viewed as an element of $G$ thanks to the compatible encoding of $G/N$):
$$\sum_{zN\in uHN/N}\mbox{\boldmath$
\ket{z^{-1}\cdot \phi_0}^{\otimes s}\ket{\phi_0}^{\otimes st}\ket{\phi_1}^{\otimes s}
\ket{\phi_1}^{\otimes st}\ket{1_G}$}
\ket{N}^{\otimes 2t}\ket{\mathsf{enc}_N(zN)},\footnote{We explicitly mention here the encoding used for the last register in order to avoid any ambiguity in the notation. Observe also that $z$ has the same encoding as $zN$ ($\mathsf{enc}_N(zN)=\mathsf{enc}(z)$).}
$$
\item Perform \OCSP$(N,\alpha,s)$
on $\ket{z^{-1}\cdot \phi_0}^{\otimes s}\ket{\phi_1}^{\otimes s}\ket{1_G}$:
{\footnotesize$$\sum_{zN\in uHN/N}\mbox{\boldmath$
\ket{z^{-1}\cdot \phi_0}^{\otimes s}\ket{\phi_0}^{\otimes st}\ket{\phi_1}^{\otimes s}
\ket{\phi_1}^{\otimes st}\ket{n_z(H\cap N)}$}
\ket{N}^{\otimes 2t}\ket{\mathsf{enc}_N(zN)}
$$}
{\rm (see the proof of Theorem~\ref{op-op} for notation and justification)}
\item Apply the group element in the last register to the first $s$ registers $\ket{z^{-1}\cdot \phi_0}$:
$$\sum_{zN\in uHN/N}\mbox{\boldmath$
\ket{\phi_0}^{\otimes s(t+1)}\ket{\phi_1}^{\otimes s(t+1)}
\ket{n_z(H\cap N)}$}
\ket{N}^{\otimes 2t}\ket{\mathsf{enc}_N(zN)}
$$
\item Left multiply by the group element in the last register 
the group element in the $2s(t+1)+1$st register
$$\sum_{zN\in uHN/N}\mbox{\boldmath$
\ket{\phi_0}^{\otimes s(t+1)}\ket{\phi_1}^{\otimes s(t+1)}
\ket{zn_z(H\cap N)}$}
\ket{N}^{\otimes 2t}\ket{\mathsf{enc}_N(zN)}
$$
\item Inverse in the last and the
$2s(t+1)+1$st  registers the computation of the compatible encoding
$\ket{zn}\ket{0}\mapsto \ket{zn}\ket{\mathsf{enc}_N(zN)}$, for every $n\in N$:
$$
\mbox{\boldmath$
\ket{\phi_0}^{\otimes s(t+1)}
\ket{\phi_1}^{\otimes s(t+1)}\ket{uH}$}
\ket{N}^{\otimes 2t}\ket{0}
$$
\end{enumerate}
\end{algo}

\begin{theorem}\label{op-op}
Let $G$ be a black-box solvable group with unique encoding of length $\enclength$
and let $N$ be 
a normal
subgroup of $G$ such that $G/N$ is abelian. 
Let $\alpha$ be a group action of $G$ and let $s,t$ be positive integers.
Then \algof{\TCS$(G,\alpha,s(t+1))$}
implements
\OCSP$(G,\alpha,s(t+1))$
using 
$(4t+1)$ implementations of 
\OCSP$(N,\alpha,s)$,
$2$ implementations of  \OCSP$(G/N,(\alpha^s)_N,t)$,
$2t$ copies of $\ket{N}$ as ancilla,
and quantum time $\mathrm{poly}(\enclength ts)$ for the remaining computation.
\end{theorem}

\begin{proof}
The complexity analysis of the algorithm is direct, only its analysis need to be detailed.
First observe that when the translating coset of $\ket{\phi_0}$ and $\ket{\phi_1}$ is empty, the
algorithm sets $H=\emptyset$ and its correctness is clear.

{From} now, we assume that  the translating coset is not empty, and it is $uH$, for some unknown $u\in G$,
where $H$ is the unknown stabilizer of $\ket{\phi_0}$ and $\ket{\phi_1}$.
Note that the translating coset of $\ket{N\cdot\phi_0}$ 
and $\ket{N\cdot\phi_1}$  for $\alpha_N$ in $G/N$ is $uHN/N$, and therefore non empty.
At Step~1, Theorem~\ref{osp-op} is applied, then
after Step~4, the algorithm has therefore computed the state $\ket{uHN/N}$ in its last register.

In Step~5, the group element in the last register is encoded using $\mathsf{enc}_N$. 
But when its inverse is applied to the first register as an element of $G$, 
we mean to use
the encoding $\mathsf{enc}$. Thanks to our definition of compatible encoding, this makes sense as long as $z$ satisfies $\mathsf{enc}_N(zN)=\mathsf{enc}(z)$. That is why the computed state becomes
a uniform superposition of states $\ket{z^{-1}\cdot\phi_0}^{\otimes s}\ldots\ket{\phi_1}^{\otimes s}\ldots\ket{\mathsf{enc}_N(zN)}$, where the superposition is over $zN\in uHN/N$, and $z$ is chosen such that $\mathsf{enc}_N(zN)=\mathsf{enc}(z)$.
For each such $z$, we prove that states $\ket{z^{-1}\cdot\phi_0}$ and $\ket{\phi_1}$
have the translating coset $n_z (H\cap N)$ over the subgroup $N$ for some $n_z\in N$
such that $\ket{n_z\cdot\phi_1}=\ket{z^{-1}\cdot\phi_0}$, meaning that
$zn_z\in uH$. 

Indeed, since $\ket{u\cdot\phi_1}=\ket{\phi_0}$, we get $\ket{(z^{-1}u)\cdot\phi_1}=\ket{z^{-1}\cdot\phi_0}$.
Therefore  $\ket{z^{-1}\cdot\phi_0}$ and $\ket{\phi_1}$ have the translating coset $z^{-1}uH$ over $G$.
Since $zN\in uHN/N$, one can write $zn_z=uh_z$, for some $h_z\in H$ and $n_z\in N$. 
Note that that both $h_z$ and $n_z$ are uniquely defined up to some element in $H\cap N$.
Then the translating coset can be rewritten as $n_zH$, implying that $\ket{z^{-1}\cdot\phi_0}$ and $\ket{\phi_1}$ 
have a non empty translating coset over $N$, which is $n_z (H\cap N)$.

Set now $H_1=\bigcup_z \big(h_z(H\cap N)\big)$.
Then after Step~9, the state of the input register is $\ket{uH_1}$.
The end of the proof consists in proving that $H_1=H$.
First observe that by definition $H_1\subseteq H$.
For the reverse inclusion, define for every $h\in H$, the coset $zN=uhN\in uHN/N$.
Chose a representative $z$ of $zN$ such that $\mathsf{enc}_N(zN)=\mathsf{enc}(z)$.
Since by construction, $zN=uhN=uh_zN$, we get
$h_z(H\cap N)=h(H\cap N)$, and therefore $h\in H_1$.
\end{proof}

If $\ket{\phi_1}=\ket{\phi_0}$ then 
$\ket{N\cdot \varphi_1}=
\ket{N\cdot \varphi_0}$ as well. Therefore the same proof shows
the following.

\begin{theorem}\label{op-op2}
Let $G$, $N$, $\alpha$, $s$ and $t$  be as in Theorem~\ref{op-op}.
Then \algof{\TCS$(G,\alpha,s(t+1))$}
implements
\SSP$(G,\alpha,s(t+1))$
using as black boxes $(4t+1)$ implementations of 
\OCSP$(N,\alpha,s)$,
$2$ implementations of  \SSP$(G/N,(\alpha^s)_N,t)$,
$2t$ copies of $\ket{N}$ as ancilla,
and quantum time $\mathrm{poly}(\enclength ts)$ for the remaining computation.
\end{theorem}

\subsection{Applications to various groups}
In this section, we study the consequences of the self-reducibility
of \OP{} for various families of solvable groups. We start by proving the
following technical statement.

\begin{theorem}
\label{op-main}
Let $G$ be a solvable
black-box group with unique encoding of length $\enclength$
and let $\alpha$ be a group action of $G$ on $\Gamma$.
Assume that we are given a subnormal series
$G=G_{0}\rhd G_1\rhd \ldots G_{r-1}\rhd G_{r}=\{1_G\}$
such that for every $1\leq i\leq r$, the factor
group is either elementary abelian of prime
exponent bounded by $e$, or $G_{i-1}/G_{i}$ 
is an abelian group of order at most $s$. Let 
$T=\Big(\big(\log\size{G}+e+2^{\sqrt{\log s}}\big)^{\Omega(e)}\log (1/\eps)\Big)^r$.
Then there exists a quantum circuit that implements
\OCSP$(G,\alpha,T)$
with error $\eps$ in quantum time $\mathrm{poly}(\ell T)$.
\end{theorem}
\begin{proof}
We actually show that, given a subnormal series
$G=G_{0}\rhd G_1\rhd \ldots G_{r-1}\rhd G_{r}=\{1_G\}$
such that for every $1\leq i\leq r$, the factor
group $G_{i-1}/G_{i}$ is either isomorphic to
$\Z_{p_i}^{n_i}$ where $p_i$ is a prime not greater than $e$
and $n_i<n$ or $G_{i-1}/G_{i}$ is an abelian group of order
at most $s$, then, 
for $T=\Big(\big(r+n+e+2^{\sqrt{\log s}}\big)^{\Omega(e)}\log (1/\eps)\Big)^r$,
there exists a quantum circuit that implements
\OCSP$(G,\alpha,T)$
with error $\eps$ in quantum time $\mathrm{poly}(\ell T)$.
{}From this the assertion follows as $n_i$ and $r$
are obviously bounded by $\log\size{G}$.

Set $u=\big(n+e+2^{\sqrt{\log s}})^{\theta(e)}$ so that for every prime 
$p\leq e$ and integer $0<n'\leq n$, for every
$\eps>0$ and for every permutation action $\alpha'$,
by Corollary~\ref{eaocsp}
\algof{\EATCS$(\Z_{p}^{n'},\alpha',
\lfloor u\log\frac{1}{\eps}\rfloor-1)$}
implements
\OCSP$(\Z_{p}^{n'},\alpha',
\lfloor u\log\frac{1}{\eps}\rfloor-1)$ with error less than $\eps/4$ and
also, for every abelian group $A$ of size at most $s$, 
\OCSP$(A,\alpha',\lfloor u\log\frac{1}{\eps}\rfloor-1)$, 
is implemented by
\algof{\AATCS$(A,\alpha',
\lfloor u\log\frac{1}{\eps}\rfloor-1)$}
(by Corollary~\ref{aaocsp})
with error at most $\eps/4$ in quantum
time less than $c_1(u\log\frac{1}{\eps})^{d_1}$.

Define
$\eps_1=\eps$ and $\eps_{j+1}=\eps_j/(9u\log\frac{1}{\eps_j})$.
Put $T'=\prod_{j=1}^r\lfloor u\log\frac{1}{\eps_j}\rfloor$. We
define a circuit by induction on $r$ that implements
\OCSP$(G,\alpha,T')$ with error at most $\eps$. In the base
case $r=1$ we use either of the two circuits discussed above.

For $r>1$, we construct the circuit by induction. Put 
$s=\prod_{j=2}^{r}\lfloor u\log \frac{1}{\eps_j}\rfloor$ and $t=\lfloor u\log\frac{1}{\eps}\rfloor-1$.
Let $N=G_1$. Then, by 
Theorem~\ref{op-op}, \algof{\TCS$(G,\alpha,s(t+1))$} implements
 \OP{$(G,\alpha,s(t+1))$} using  $(4t+1)$ implementations of
\OP{$(N,\alpha,s)$}, $2$ implementations of \OP{$(G/N,(\alpha^{s})_N,t)$},
$2t$ copies of $\ket{N_{r-1}}$.
and quantum time less than $c_2(\ell st)^{d_2}$. 

By the assumption on $u$, \OP{$(G/N,(\alpha^{s})_N,t)$} can be implemented by
with error less than $\eps/4$ in quantum time less than
$c_1(\ell ts)^{d_1}$.
(The oracle for $\alpha^s$ is implemented by $s$ applications 
of the oracle for $\alpha$.)
Now, by induction \OP{$(N,\alpha,s)$} can be implemented with error 
$\eps/9t<\eps/(8t+2)$
in quantum time $c(\ell s)^{d}$, using $\Order(s)$ copies of $\ket{N_i}$, 
for $1\leq i\leq r-2$. The overall error is clearly less than $\eps$.

We show that $T'=\big((ru)^{O(1)}\log\frac{1}{\eps}\big)^r$.
To see this, observe that
$\log\frac{1}{\eps_{j+1}}=\log\frac{1}{\eps_{j}}+\log 9u+
\log\log\frac{1}{\eps_{j}}$. By induction on $j$,
we can show that 
$\log\frac{1}{\eps_{j}}\leq j^2u\log\frac{1}{\eps}$,
if $u$ is large than an appropriate constant.
(Indeed, the induction hypothesis gives
$\log\frac{1}{\eps_{j+1}}\leq j^2u\log\frac{1}{\eps}
+\log 9u+2\log j+\log\log\frac{1}{\eps}\leq
(j+1)^2u\log\frac{1}{\eps}$ if $u$ is sufficiently large.)
Therefore
$T'\leq u^r\prod_{j=1}^r\log\frac{1}{\eps_j}\leq
(ru)^{2r}\big(\frac{1}{\eps}\big)^r$.

The quantum time is bounded by 
$c_2(\ell st)^{d_2}+c_1(\ell st)^{d_1})+c(4t+2)s^d<
c{\ell T'}^d$ if $c$ and $d$ are sufficiently large.
\end{proof}

The theorem above gives a polynomial time algorithm
for $\OP$ in abelian groups of constant exponent. 
More generally, we have the following.
\begin{theorem}\label{smooth-abelian}
Let $G$ be an abelian black-box group with unique encoding of
length $\enclength$
and let $\alpha$ be a group action of $G$ on $\Gamma$.
Assume that $G$ has a subgroup $N$
of  exponent at most $e$ such that $G/N$ has size
an most $s$.
Let $T=\big((\log \size{G}+e+2^{\sqrt {\log s}})^{\Omega(e)} \log(1/\eps)\big)^{\log e}$.
Then there exists a quantum circuit that implements
\OCSP$(G,\alpha,T)$
with error $\eps$ in quantum time $\mathrm{poly}(\ell T)$.
\end{theorem}
\begin{proof}
Using for instance~\cite{mos04},
a decomposition of $G$ as a direct product of
cyclic subgroups $H_1,\ldots,H_m$ of prime power
order $p_i^{\alpha_i}$ can be can be computed in quantum time 
$\mathrm{poly}(\ell)$. Considering only
indices $i$ such that $p_i\leq e$, by an 
exhaustive search we can find
in time polynomial in $\log\size{G}^{O(e)}$ 
integers $\beta_i\leq \alpha_i$ ($i=1,\ldots,m$) subject
to the constraint $\mbox{lcm}\{p_i^{\beta_i}|i=1,\ldots,m\}\leq e$
such that $\prod_{i=1}^mp_i^{\beta_i}$ is maximal. Then the
sum $G_1$ of the subgroups $H_i^{p_i^{\alpha_i-\beta_i}}$ is
the largest (by cardinality)
subgroup of exponent at most $e$, consequently $\size{G/G_1}\leq s$.
Let $e=q_2\cdots q_r$ where $q_i$ are not necessarily distinct primes
and let $G_{i+1}=G_{i}^{q_i}$ ($i=2,\ldots,r$). Then $r\leq \log e$
and we can apply Theorem~\ref{op-main} to the sequence
$G>G_1>\ldots>G_r$. 
\end{proof}

Using a similar proof we obtain the following generalization.

\begin{theorem}\label{smoothly-solvable}
Let $G$ be a solvable
black-box group with unique encoding of length $\enclength$
and let $\alpha$ be a group action of $G$ on $\Gamma$.
Assume that $G$ has derived length $m$ and
that for every index $0<i\leq m$,
the factor of the subsequent derived 
subgroups ${\widetilde {G_{i-1}}}=G^{(i-1)}/G^{(i)}$
have a subgroup ${\widetilde {N_{i-1}}}$ of exponent at most $e$
such that $\size{{\widetilde {N_{i-1}}}/{\widetilde {N_{i-1}}}}\leq s$.
Let $T=\Big(s\big((\log\size{G}+e)^{\Omega(e)}\log (1/\eps)\big)^{\log e}\Big)^m$.
Then there exists a quantum circuit that implements
\OCSP$(G,\alpha,T)$
with error $\eps$ in quantum time $\mathrm{poly}(\ell T)$.
\end{theorem}

The following theorem describes the class of groups
for which our methods give polynomial time hidden subgroup
algorithms. Recall that a smoothly
solvable group has constant derived length and the factors 
${\widetilde {G_{i-1}}}=G^{(i-1)}/G^{(i)}$ satisfy the condition
of the preceding corollary with constant $e$ and $s=\poly(\size\log{G})$.

\begin{theorem}\label{htp-hsp-thm}
\OP and \HTP{} can be solved over smoothly solvable groups
in quantum polynomial time. Furthermore, \SP{} and \HSP{} 
can be solved over solvable groups having a smoothly solvable 
commutator subgroup in quantum polynomial time.
\end{theorem}
\begin{proof}
The first statement follows directly from the preceding theorem, using
Proposition~\ref{htp-op}.
For the second 
part we additionally use Theorem~\ref{op-op2}.
\end{proof}

By \cite{glasby89}, every solvable group has derived series of
length $m=\Order(\log\log\size{G})$. Using this result and
Theorem~\ref{smoothly-solvable},
we get a 
quasi-polynomial 
quantum algorithm for all solvable groups
of constant exponent.
\begin{theorem}\label{loglog-thm}
Let $G$ be a 
solvable black-box group with unique encoding of length $\enclength$ 
and of constant exponent.
Then \HTP{}$(G)$ can be solved 
in quantum time 
$\enclength^{\Order(1)} (\log\size{G})^{\Order(\log\log\size{G})}$.
Furthermore, the \HSP{} can be solved 
in quantum time 
$\enclength^{\Order(1)} (\log\size{G})^{\Order(\log\log\size{G})}$
in groups $G$ for which $G'$ has constant exponent. 
\end{theorem}

Finally, an application of Theorem~\ref{op-main} with $e=1$ and $s=\size{G}$
to the derived series of a solvable group gives the following.

\begin{theorem}
\label{kup-thm}
Let $G$ be a 
solvable black-box group with unique encoding of length $\enclength$.
Then \HSP{}$(G)$ and \HTP{}$(G)$ can be solved with constant error
in quantum time 
$\enclength^{\Order(1)} (\log\size{G})^{\Order(\sqrt{\log\size{G}} \cdot \log\log\size{G})}$. 
\end{theorem}

\section*{Acknowledgements}
We wish to thank Mark Ettinger and Peter H\o{}yer
for sharing their knowledge and ideas about \HTP{} with us,
and Martin R\"otteler for several useful discussions on \HSP{}.
We are also grateful to Cris Moore, Csaba Schneider and Yong Zhang
for helpful discussions regarding a previous version of this article,
and for pointing out reference~\cite{glasby89}.

\end{document}